# Label-Free Whole Slide Virtual Multi-Staining Using Dual-Excitation Photon Absorption Remote Sensing Microscopy

James E.D. Tweel[1], Benjamin R. Ecclestone[1], James A. Tummon Simmons[1], and Parsin Haji Reza[1,*]

[1]*PhotoMedicine Labs, Systems Design Engineering, University of Waterloo, Waterloo, Ontario, N2L 3G1, Canada,*
∗ *Corresponding author:* `phajireza@uwaterloo.ca`

**Abstract** – Histochemical staining is essential for visualizing tissue architecture and cellular morphology but is destructive and limited by the availability of tissue for multiple stains. Virtual staining with label-free microscopy offers a non-destructive alternative, enabling multiple stains to be generated from the same section while reducing stain variability and preserving tissue for downstream assays. Here, a new dual-excitation Photon Absorption Remote Sensing (PARS) system is presented, representing the first application of long-wave ultraviolet A (UVA) 355 nm excitation alongside the established UVC 266 nm source. The addition of 355 nm extends PARS contrast beyond the nuclear and connective tissue signals of 266 nm, enhancing visualization of stromal architecture (e.g., collagen, elastin) and capturing red blood cells, melanin-containing elements, and other tissue features through complementary radiative and non-radiative absorption. The 266 nm and 355 nm pulses interrogate the sample in an interlaced fashion, enabling concurrent acquisition without compromising imaging speed. Using the RegGAN image-translation framework, this work presents the first demonstration of PARS virtual staining across multiple specialized stains, including Masson's trichrome, Periodic acid–Schiff (PAS), and Jones methenamine silver (JMS), in addition to hematoxylin and eosin (H&E), across diverse human and murine tissues. A masked evaluation by expert pathologists showed that virtual stains achieved the same diagnostic quality as their chemical counterparts, and pathologists could not reliably distinguish real from virtual stains. By providing label-free multi-stain outputs from a single scan, dual-excitation PARS virtual staining could integrate into digital pathology workflows, expanding diagnostic utility. Real and virtual whole-slide image (WSI) pairs are publicly available at the BioImage Archive (https://doi.org/10.6019/S-BIAD2232).

## 1. Introduction

Histochemical stains are the primary method for visualizing the microanatomy of tissue and remain essential in disease diagnostics and life-science research [1]. By binding to specific biochemical structures, stains impart contrast in otherwise transparent tissue sections. The stained samples are examined using light microscopy, often with brightfield whole-slide scanners to capture high-resolution, sub-cellular detail across the sample for digital viewing, navigation, and computational analysis [2], [3]. The most widely used stain is hematoxylin and eosin (H&E), which stains nuclei purple and cytoplasm and extracellular components pink, and is routinely used to assess tissue architecture and cellular morphology, particularly in cancer diagnosis [4]. In addition, specialized stains provide more targeted contrast, such as Masson's trichrome for collagen analysis, Periodic acid-Schiff (PAS) for carbohydrate-rich structures, and Jones methenamine silver (JMS) stain for highlighting glomerular and tubular membranes in renal pathology [1], [4].

While invaluable, histochemical staining has many drawbacks. The process is destructive, and the same section cannot always be re-stained or used for ancillary diagnostic assays [5]. Multiple stains require additional cutting and consumption of limited biopsy tissue, and adjacent cuts inevitably have misalignments or mismatched structures. Beyond tissue loss, staining protocols are complex, multi-step procedures that require trained histotechnologists and dedicated laboratory infrastructure, and stain quality can be highly variable [6]. These workflows also generate chemical waste, add cost, and, when additional stains are required, can lengthen turnaround times and delay diagnosis and treatment [7].

To address these challenges and complement traditional staining, a range of label-free microscopes have been developed that capture endogenous tissue contrast without the need for chemical markers. These methods leverage light-matter interactions such as scattering (e.g., quantitative phase imaging [8], QPI) and electronic absorption via radiative and non-radiative pathways (e.g., autofluorescence [9], photoacoustic [10], and photothermal microscopy [11]), along with non-linear counterparts including multiphoton fluorescence [12], harmonic generation [13], and stimulated Raman scattering (SRS) microscopy [14]. Advances in computational methods have extended the utility of label-free imaging modalities in histology through virtual staining, which uses deep learning to translate raw images into familiar brightfield or fluorescent images of standard histochemical or immunohistochemical stains [7]. This enables multiple stains to be visualized on the same section, without the time, cost, or tissue loss associated with chemical staining, while preserving samples for downstream histological or molecular analyses.

A variety of label-free modalities have been explored for deep learning-based virtual staining. Autofluorescence imaging has been the most widely explored, having shown virtual H&E and other histochemical stains [15], [16], [17], [18], [19], [20], as well as immunohistochemical stains [21]. Fluorescence lifetime imaging (FLIM) has similarly been applied to virtual H&E staining [22], [23], while scattering-based methods such as QPI [24], [25] and brightfield imaging [26], [27] have been used to generate multiple virtual stains. Non-radiative absorption techniques, including ultraviolet (UV) photoacoustic-based approaches have been used for virtual H&E staining [28], [29], [30], [31]. Nonlinear techniques, including stimulated Raman scattering (SRS) [32], [33] and multimodal nonlinear methods [34], have also demonstrated the ability to generate H&E-like virtual stains.

The performance of virtual staining depends on the endogenous contrasts captured by the label-free modality. Input that provides greater biochemical specificity or complementary contrast enables the model to better distinguish tissue components and reconstruct the features of the target stain [15], [22], [35]. For this reason, Photon Absorption Remote Sensing (PARS) microscopy is well suited for virtual staining because it simultaneously captures radiative and non-radiative absorption signals from a single excitation event, effectively combining the contrast information typically provided by autofluorescence (radiative pathway) and photoacoustic or photothermal methods (non-radiative pathway). Uniquely, the PARS signals are intrinsically linked to the same absorption event and therefore encode the total absorption of the target and the ratio of relaxation fractions (quantum efficiency ratio, QER [36], [37]). For label-free histology, PARS has primarily been implemented with a deep UV 266 nm excitation source. With this wavelength, non-radiative (NR) absorption provides important nuclear contrast, while radiative (R) autofluorescence highlights cytoplasmic and extracellular matrix (ECM) structures, together producing contrast highly analogous to H&E [36], [38]. In this configuration, PARS has previously been used for virtual H&E staining using both paired [39], [40] and unpaired training schemes [41], [42], and has demonstrated high diagnostic concordance in blinded studies of cancerous human breast [42] and skin tissues [40].

While previous PARS histology studies have focused exclusively on H&E emulation, this work presents the first demonstration of PARS virtual staining across multiple specialized stains, including Masson's trichrome, PAS, and JMS, in addition to H&E. This advance is achieved with a newly developed dual-excitation PARS system, representing the first application of long-wave UVA 355 nm excitation alongside the established 266 nm UVC source. The addition of 355 nm extends PARS contrast by targeting the absorption of multiple histologically relevant chromophores, providing complementary radiative and non-radiative contrasts to 266 nm [9], [10], [43], [44]. Radiative signals provide visualization of stromal architecture via collagen and elastin autofluorescence [43], while non-radiative absorption highlights red blood cells (RBCs) through hemoglobin absorption and reveals melanin-containing structures such as melanocytes [10].

Virtual staining is demonstrated in unseen whole slide images (WSI) across a diverse set of tissues and disease contexts, including human kidney with clear cell renal cell carcinoma (ccRCC), human skin with nodular melanoma and fungal infection, and multiple murine organs such as kidney, gastrointestinal tract (GI), and brain, using large datasets that enable robust evaluation in both healthy and pathological conditions. An improved virtual staining model, RegGAN [45], is used. This supervised framework integrates a registration network, combining the accuracy of paired approaches with the misalignment tolerance of unpaired methods, and is shown to outperform Pix2Pix and CycleGAN across all tissue types and disease cases. Furthermore, results demonstrate that the dual-excitation input consistently outperforms 266 nm or 355 nm alone, with quantitative improvements in similarity and perceptual metrics and improved reconstruction of histology-specific features, such as nuclei, collagen, RBCs, and fungal hyphae. Importantly, masked evaluation by three expert pathologists confirmed that virtual stains achieve comparable diagnostic quality to their chemical counterparts and that pathologists were unable to reliably determine whether a given image was chemically or virtually stained. For each tissue–stain combination, corresponding chemical and virtual WSIs are publicly available at the BioImage Archive [46] (see Data Availability).

## 2. Methods

### 2.1 Dual-Excitation PARS System Architecture and Configuration

The dual-excitation PARS histology microscope system used in this study is illustrated in Figure 1a. The system includes two excitation sources: a 400ps pulsed 266 nm UVC laser (Wedge XF 266, Bright Solutions) and a 1.5 ns pulsed 355 nm UVA laser (ULPN-355-10-1-10-M, IPG Photonics), both operating at a 50kHz pulse repetition rate. These excitation wavelengths were selected to provide complementary contrast for virtual staining. As shown in previous works [36], [38], 266 nm offers strong absorption of nucleic acids, providing important nuclear-specific non-radiative contrast, along with radiative emissions from ECM structures. The 355 nm source is similarly broadly absorbed, providing complementary autofluorescence from ECM components such as collagen and elastin [43], along with non-radiative signals from absorbers like hemoglobin for RBC visualization, melanin, and other pigments [10].

The output from the 266 nm source first passes through a $CaF_2$ prism (PS862, Thorlabs) to spectrally separate residual 532 nm green light, which is directed to a beam trap (BT: BT610, Thorlabs). The 266 nm beam is then 3× expanded using a variable beam expander (3×VBE: BE03-266, Thorlabs). The adjustable collar of the expander allows for collimation control of the 266 nm laser for axial beam alignment. The expanded beam is directed toward a Nd:YAG harmonic separator ($HS_2$: 37-721, Edmund Optics) where it is combined with the 355 nm source.

The output from the 355 nm source is first condensed by 2× and then re-expanded with a 2× variable expander (2×VBE: BE02-355, Thorlabs), resulting in an overall magnification of 1X. As with the 266 nm, this is done so the adjustable collar of the expander can be used to fine-tune the collimation of the 355 nm laser for axial alignment purposes. The 355 nm beam is then reflected off a second Nd:YAG harmonic separator ($HS_1$: 37-721, Edmund Optics) and transmitted through $HS_2$ to align with the 266 nm path.

A continuous wave 405 nm detection source (OBIS-LS 405, Coherent) is used to probe the sample for scattering signals and non-radiative contrast. The fiber-coupled laser is first collimated (Col.: C40APC-A, Thorlabs) and directed through both $HS_1$ and $HS_2$ to co-align with the two excitation beams.

All three beams (266 nm, 355 nm, 405 nm) are then co-focused onto the sample using a 0.42NA UV objective lens ($OL_1$: NPAL-50-UV-YSTF, OptoSigma). The forward-propagating light, including scattered 405 nm and 355 nm and radiative emission from both UV sources, is collected using a 0.7 NA top objective lens ($OL_2$: 278-806-3, Mitutoyo). The collected light is coupled into a multimode optical fiber (MMF: M133L01, Thorlabs) using a high-NA achromatic coupler ($AC_1$: F950FC-A, Thorlabs), and then re-collimated into free space by $AC_2$ for downstream spectral separation.

The scattered 355 nm light is immediately sent into a beam trap using a 355 nm notch filter ($NF_1$: 39-387, Edmund Optics). The transmitted 405 nm beam is then separated from the radiative emissions using a 405 nm notch filter (NF$_2$: NF405-13, Thorlabs). The radiative emissions and non-radiative scattering modulation are each recorded on avalanche photodetectors (PD: APD130A2, Thorlabs) following condenser lenses (Cond.: ACL25416U-A, Thorlabs). An additional avalanche photodetector is used in the forward path to record the 266 nm pulse amplitude and 405 nm power level; the 355 nm source was not recorded given its lower pulse variability. These beams are sampled using beam samplers (BS1: BSF10-UV and BS2: BSF10-A, Thorlabs) for post-acquisition power correction.

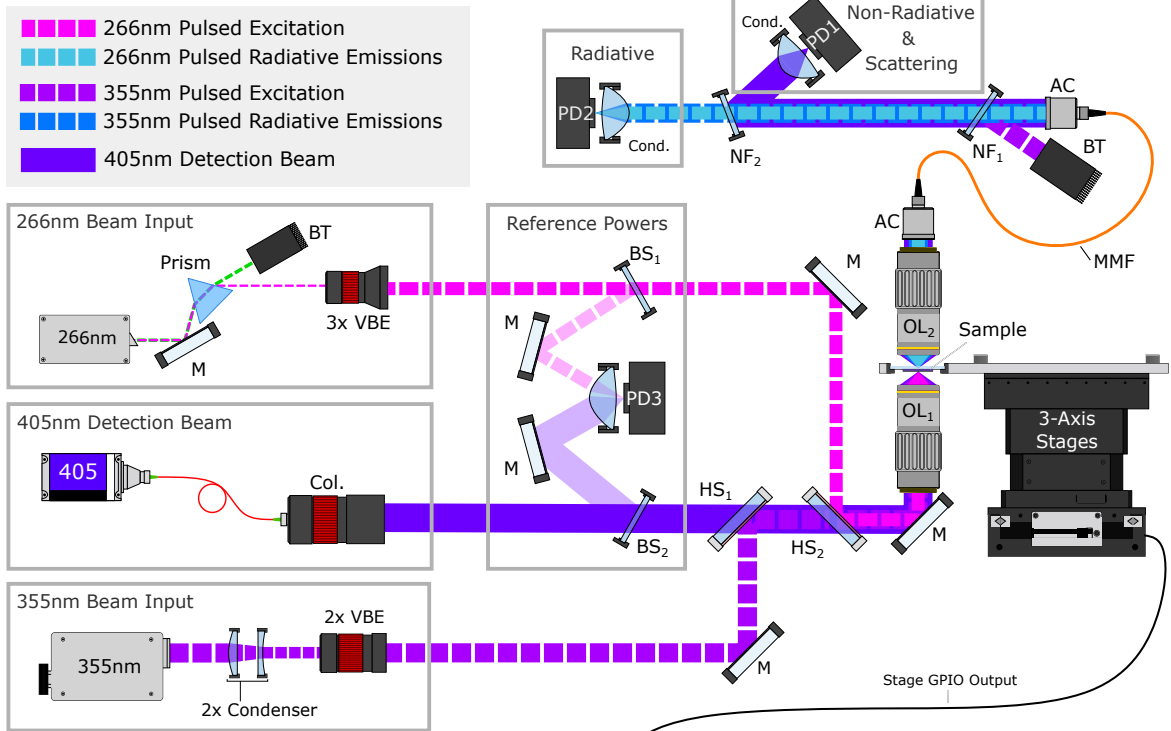
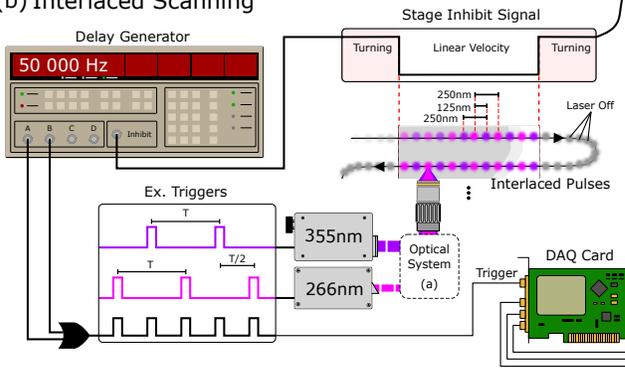
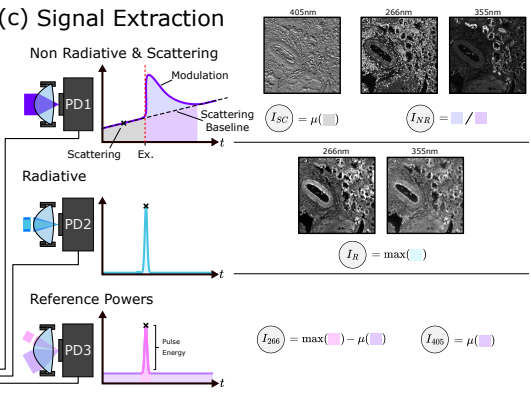

**Figure 1:** Dual-excitation PARS architecture, interlaced scanning pattern, and signal extraction. (a) Overview of the PARS microscope optical system. Component labels: AC (achromatic collimator), BT (beam trap), BS (beam sampler), Col. (collimator), HS (harmonic separator), MMF (multi-mode fiber), M (mirror), NF (notch filter), OL (objective lens), PD (photodiode detector), VBE (variable beam expander), Cond. (condenser). (b) Interlaced excitation scan pattern. A delay generator alternately triggers 266 nm and 355 nm pulses to create interlaced excitation pulses on the sample. Pulses are inhibited during stage turnaround via a programmed stage GPIO signal. A high-speed digital acquisition (DAQ) card is triggered at each excitation pulse to collect time-series data at each event. (c) Signal extraction from system photodetectors. At each pulse, ∼500 ns segments are recorded from three photodetectors and compressed into pixel values for each contrast. This produces non-radiative and radiative images for each excitation wavelength, and a 405 nm scattering image. Non-radiative contrast is calculated as the modulation from the scattering baseline; radiative signal is defined as the peak value; and reference powers (266 nm pulse amplitude and 405 nm power) are extracted for post-acquisition power correction.

## 2.2 Interlaced Scanning, Signal Acquisition, and Image Reconstruction

An overview of the scanning process, signal acquisition, and image reconstruction pipeline is illustrated in Figure 1b and 1c. A programmable delay generator (DG645, Stanford Research Systems) is used to alternately trigger the 266 nm and 355 nm lasers, each at 50 kHz, producing spatially interlaced excitation spots along the scan trajectory. By interlacing the pulses rather than performing two sequential scans, the dual-excitation system acquires radiative and non-radiative contrast from both excitation sources essentially simultaneously, without compromising imaging speed.

During a scan, the mechanical stages move the samples in an S-shaped raster pattern. In the linear velocity regions, the stages move at 12.5 mm/s, spacing the alternating 266 nm and 355 nm pulses by 125 nm, and maintaining a 250 nm step between successive pulses of the same wavelength.

During the turnaround (non-linear) regions, both excitation lasers are disabled to avoid high-density UV pulses and potential photobleaching. A TTL high signal from the stage is used to inhibit the delay generator, effectively gating off the laser triggers during these periods.

At each excitation pulse, a high-speed digitizer (Gage CSE1442, 200 MS/s) captures a ~500 ns time segment from each system photodetector, as shown in Figure 1c. In total, three photodetectors (PDs) are used to collect signals that produce a total of five distinct PARS contrasts in a single pass: 405 nm scattering, 266 nm and 355 nm non-radiative (PD1), and 266 nm and 355 nm radiative, along with 266 nm and 405 nm input reference powers (PD3).

PD1 captures the forward-scattered 405 nm signal and non-radiative response. The scattering contrast is calculated as the average pre-excitation scattering intensity. These scattering images are extracted during the autofocus phase of the whole-slide scan to ensure high-SNR acquisition. Non-radiative contrast is computed from the transient modulation of the scattering signal relative to a predicted baseline following excitation. As described in [47], the scattering baseline is estimated using a first-order linear fit of the scattering signal before and after excitation. The non-radiative signal is then calculated as the percentage deviation from this predicted baseline. This approach has been shown to improve SNR [47], particularly at structural edges where scattering intensity varies rapidly, by reducing error from motion-induced baseline shifts.

PD2 captures the radiative emission signals from both excitation wavelengths. The radiative contrast is extracted by taking the peak signal amplitude within the acquisition window. PD3 measures the 266 nm and 405 nm input reference powers. The average scattering input power is estimated from the pre-excitation signal level, while the 266 nm pulse energy is calculated as the difference between the peak signal amplitude and the scattering power reference.

Following the extraction of pixel intensities for each excitation pulse, the 266 nm and 355 nm pixel values are deinterlaced and reconstructed into separate images by mapping each pixel onto a cartesian grid based on their stage positions. Each reconstructed image represents a single 500 µm x 500 µm field of view. Using the automated workflow detailed in [38], whole slide imaging is performed by autofocusing across the sample using the scattering signal and capturing a series of overlapping sections. These sections are then stitched and blended to form a whole slide image of each contrast.

## 2.3 Deep Learning Based Virtual Staining

*Whole Slide Preprocessing, Artifact Annotation, and Whole Slide Inference*

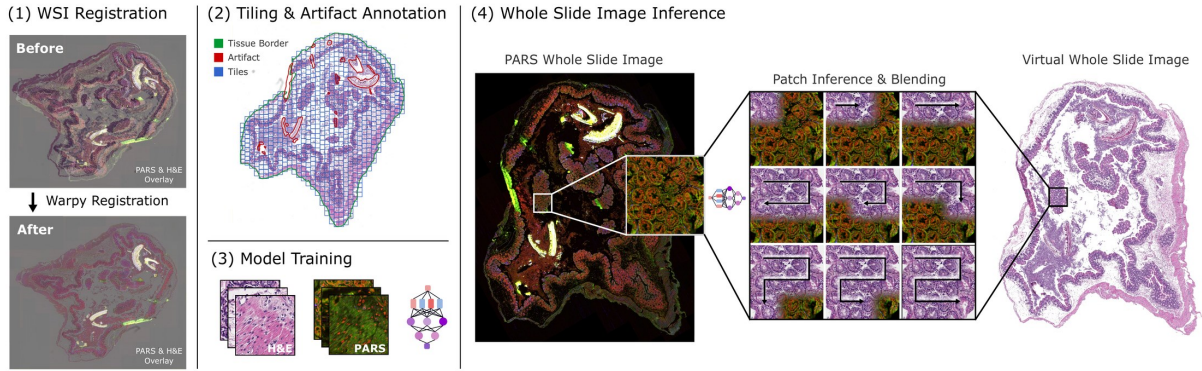

**Figure 2:** Workflow for preparing training data and performing whole slide virtual staining. (1) PARS and stained WSIs are aligned using the Warpy registration framework. (2) Tissue borders and artifacts are annotated in QuPath, and the slides are tiled into artifact-free patches. (3) These patches are used to train the virtual staining model. (4) Virtual staining is applied to full PARS WSIs using overlapping patch-wise inference, followed by linear blending and tissue masking to produce the final virtual stained slide. The example shown depicts mouse H&E gastrointestinal tissue.

The process used to prepare the training data for the virtual staining model and to perform WSI inference is illustrated in Figure 2. Following PARS imaging, the same unstained tissue sections were subsequently stained and brightfield scanned. Due to the differences in the imaging modalities and tissue deformation during the staining, spatial misalignments between the PARS and stained images were then corrected using the Warpy registration pipeline [48]. Warpy applies a combination of rigid and non-rigid transformations to closely align the two WSI pairs, a necessary starting point for the model training.

After registration, QuPath was used to annotate the tissue borders and label artifacts across the tissue. These artifacts include tissue tears, cover-slip bubbles, folds, and out-of-focus areas in the stained images, as well as paraffin obstructions and dust artifacts in the PARS images. Annotations were exported in GeoJSON format and processed by a custom Python script, which excluded the artifact regions and tiled the WSIs into smaller, co-registered patches. Tiling was required to divide the high-resolution slides into manageable chunks for training.

The resulting datasets were used to train the virtual staining model, described in the following section. Once trained, the model was applied to the entire PARS WSIs using a patch-based inference with 40% overlap between adjacent tiles. Overlapping patch regions were linearly blended to ensure visual continuity, resulting in a virtual WSI resembling a conventional stained slide. Following inference, a tissue mask was applied to remove any background artifacts. The mask was generated by thresholding the 355 nm radiative contrast channel from the PARS image, and the background regions were smoothly faded to white.

*Model Architecture and Training Scheme*
Virtual staining employs image-to-image translation models, most of which follow one of two training paradigms: supervised approaches like Pix2Pix and variants [49], [50], and unsupervised methods such as CycleGAN [51]. Both approaches have been widely used in virtual staining, with unpaired methods more common due to general scarcity and difficulty in obtaining perfectly registered training data [52].

Supervised models like Pix2Pix are effective when training pairs are precisely aligned, however, achieving such alignment is particularly difficult in high-resolution, cross-modality tasks like PARS to brightfield translation. Even with registration pipelines like Warpy, small residual misalignments exist due to differences in the optical systems, imaging conditions, and local tissue deformation during staining. These inconsistencies introduce label "noise" that can result in blurred or smeared output quality. Pix2Pix has previously been applied to PARS data for H&E emulation [39], [40], demonstrating strong performance and high concordance under carefully registered conditions; however, at the cost of labor-intensive preprocessing and time-consuming annotation to identify and exclude poorly aligned regions, which also reduced the effective dataset size.

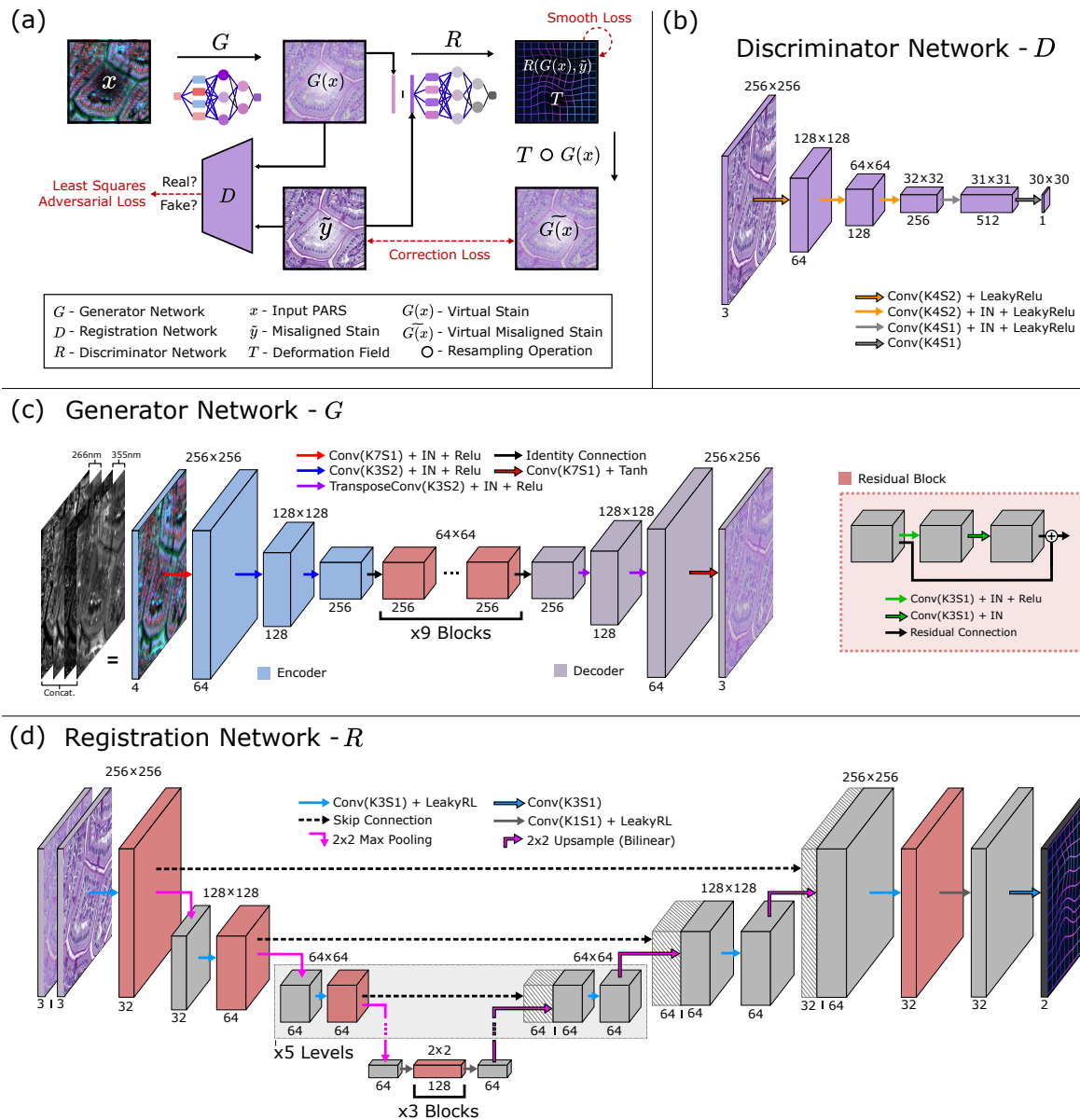

**Figure 3:** Overview of the RegGAN-based virtual staining framework and network architectures. (a) Schematic of the RegGAN training workflow. The generator network $G$ receives a multichannel PARS input $x$ and produces a virtual stain $G(x)$. This is concatenated with the misaligned real stain $\tilde{y}$ and passed through the registration network $R$, which predicts a deformation field $T$ that is applied to align the virtual stain. A correction loss is computed between the registered output and the real stain, while a smoothness loss is applied to $T$ to enforce spatial smoothness. An adversarial discriminator $D$ guides $G$ via a least squares loss. (b) Architecture of the discriminator $D$, based on a 70×70 PatchGAN, which outputs a patch-wise classification map for real/fake discrimination. (c) Architecture of the generator $G$, a ResNet-based encoder-decoder that processes four-channel 266 nm and 355 nm contrast images and outputs an RGB virtual stain. (d) Architecture of the registration network $R$, a ResUNet-style model that predicts a 2D deformation field from a six-channel concatenation of the real and virtual stain. IN = Instance Normalization; ReLU = Rectified Linear Unit.

CycleGAN-based models avoid the alignment issue by eliminating the need for paired training data. However, the lack of pixel-level supervision results in weaker constraints on both contrast translation and structural preservation. CycleGAN has previously been applied to loosely paired PARS data for virtual H&E staining [41], successfully avoiding the blurring observed with Pix2Pix under misalignment. While this worked well for PARS to H&E, where the mapping between modalities was more direct, the stains explored in this work involve more complex or subtle structural and contrast relationships, making a supervised approach better suited to learn these mappings.

For these reasons, the virtual staining model used in this work is based on RegGAN [45], which offers a more effective balance between the paired and unpaired training regimes. RegGAN treats misalignments in the target image as a form of noisy labelling and incorporates a learnable registration network into the training process. This network adaptively estimates deformation fields that align the generated image to the target domain. As such, the model retains the benefits of paired supervision while remaining robust to residual misalignment, making it well suited for high-resolution virtual staining across imaging modalities.

The RegGAN training scheme is illustrated in Figure 3a. The generator network ($G$) receives a multichannel PARS input patch ($x$) and produces a virtual stained output ($G(x)$). This output, along with its corresponding misaligned real stain image ($\tilde{y}$), is concatenated and passed through the registration network ($R$). The registration network predicts a deformation field ($T$), which is applied to $G(x)$ to generate a registered virtual output $\widetilde{G(x)}$ that is more closely aligned with the real stain. An L1 correction loss is computed between the registered virtual output and the real stain image to encourage structural and visual similarity in the generated output.

In parallel, the discriminator ($D$) learns to distinguish between real and virtual images, while the generator works to fool the discriminator by producing increasingly realistic virtual stains. To improve training stability, a least squares adversarial loss is used, per the LSGAN framework [53]. Additionally, a smoothness loss is applied to the predicted deformation field, encouraging spatial smoothness by penalizing large spatial gradients.

Figure 3(b-d) detail the convolutional architectures of the discriminator ($D$), generator ($G$), and registration ($R$) networks, respectively. Full layer configurations, including convolutional kernel sizes, strides, and feature dimensions, are provided within the figure. The discriminator network, shown in Figure 3b, follows the 70x70 PatchGAN architecture which outputs a patch-wise classification map, where each value reflects the likelihood of a local region being real or generated. The localized feedback helps the generator in producing realistic fine-scale texture and structural detail.

The generator network, illustrated in Figure 3c, adopts a ResNet-based encoder-decoder structure, commonly used in image-to-image translation tasks [51]. The inputs to the generator are the concatenated non-radiative and radiative contrasts from both the 266 nm and 355 nm wavelengths. The network consists of an encoder with two downsampling layers, a bottleneck with nine residual blocks, and a decoder that upsamples the features and outputs a three-channel RGB virtual stain image.

The registration network, shown in Figure 3d, is a ResUNet-style architecture from [54]. It receives a six-channel input formed by concatenating the virtual and real stain images. The architecture consists of seven downsampling levels, a central bottleneck with three residual blocks, and a symmetric decoder with bilinear upsampling and skip connections from the encoder. The final output is a two-channel deformation field representing the predicted $x$ and $y$ displacements.

*Training Parameters and Implementation Details*

An overview of the tissue types, stains, and dataset sizes used in this study is provided in Table 1. The datasets include mouse gastrointestinal (GI), kidney, and brain tissues; human kidney samples with ccRCC; human skin with fungal infection; and human skin with nodular melanoma. The stains used include hematoxylin and eosin (H&E), Masson's trichrome, PAS, and JMS.

For each tissue and stain combination, the dataset was split into 80% training and 20% validation patches. In addition, multiple WSIs were held out entirely for testing for each virtual staining model. WSI-based evaluation provided a realistic and stringent test scenario, as these slides were never seen during training or validation. For each WSI, a global mean and standard deviation were computed for each contrast channel, and each PARS input patch was normalized by subtracting this mean and dividing by the standard deviation.

**Table 1:** Training and validation sample summary by tissue type and stain.

| Tissue Type | Histochemical Stain | Training Samples |
| --- | --- | --- |
| Human kidney (ccRCC) | Hematoxylin and eosin | 72,749 |
| Human skin (melanoma) | Hematoxylin and eosin | 26,066 |
| Human skin (fungal disease) | Periodic acid–Schiff | 21,030 |
| Human kidney (ccRCC) | Masson's trichrome | 88,769 |
| Mouse kidney | Jones methenamine silver | 27,157 |
| Mouse kidney | Hematoxylin and eosin | 16,524 |
| Mouse kidney | Periodic acid–Schiff | 22,586 |
| Mouse kidney | Masson's trichrome | 28,117 |
| Mouse gastrointestinal | Hematoxylin and eosin | 9,481 |
| Mouse gastrointestinal | Periodic acid–Schiff | 29,568 |
| Mouse gastrointestinal | Masson's trichrome | 33,425 |
| Mouse brain | Hematoxylin and eosin | 21,103 |

*Note:* Samples are 256×256px patches.

All models were trained using the Adam optimizer with betas (0.5, 0.999). The generator and registration networks used a learning rate of $2\times10^{-4}$, while the discriminator used a lower learning rate of $2\times10^{-5}$. The generator was set to update four times for every update of the discriminator and registration networks. A batch size of 1 and a fixed patch size of 256×256 pixels was used. All LeakyReLU activations had a negative slope of 0.2. The smooth loss, correction loss, and adversarial loss were weighted by 10, 20, and 1, respectively. Data augmentation included random horizontal and vertical flips, along with 90° rotations.

Training was conducted for up to 70 epochs or until validation losses plateaued. The discriminator was updated using a history buffer containing the 50 most recent generated images, rather than relying solely on the latest outputs, as suggested in [55]. All models were implemented in Python 3.11.9 using PyTorch 2.4.1+cu118, and training was performed on a single NVIDIA RTX 4070 Super GPU using mixed-precision computation.

In addition to the primary models with all contrasts, ablation experiments were conducted to evaluate how different subsets of PARS contrast channels affect virtual staining performance. Separate models were trained using only 266 nm radiative and non-radiative inputs, as well as only 355 nm radiative and non-radiative inputs. All training parameters remained consistent with the primary models. Some contrast configurations had more limited input information, and the generator update frequency was initially increased during the first ~10 epochs to improve early training stability, after which the standard 4:1 generator to discriminator and registration network update ratio was resumed.

## 2.4 Sample Preparation

Tissue samples were sectioned at ~5$\mu$m thickness onto microscope slides from formalin-fixed paraffin-embedded (FFPE) blocks prepared using standard histological methods. Freshly resected tissues were fixed in formalin within 20 minutes of excision and stored for 24 to 48 hours. Samples were then dehydrated in ethanol, cleared with xylene, and embedded in liquid paraffin wax at 60°C. Slides were baked at a low temperature (50–55 ºC) to smooth the paraffin surface for PARS imaging.

After PARS imaging, tissues were stained with one of the following: hematoxylin and eosin (H&E), Masson's trichrome, PAS, or JMS, and scanned at 40× magnification using a brightfield microscope (Leica Aperio AT2, Robarts Research Institute).

The ccRCC human kidney samples were provided by the Ontario Tumour Bank. Human skin tissues (nodular melanoma and fungal infection) were provided by collaborators at the Cross Cancer Institute (Edmonton, Alberta, Canada). Murine tissues were resected in-house and sent to the Robarts Research Institute (University of Western Ontario) for FFPE processing.

Human samples were from anonymous patient donors and no information regarding patient identity was provided to the researchers. Patient consent was waived by the ethics committee because these archival tissues were no longer required for patient diagnostics. Samples were collected under protocols approved by the Research Ethics Board of Alberta (Protocol ID: HREBA.CC-18- 0277) and the University of Waterloo Health Research Ethics Committee (Photoacoustic Remote Sensing (PARS) Microscopy of Surgical Resection, Needle Biopsy, and Pathology Specimens; Protocol ID: 40275). All human tissue experiments were conducted in accordance with the government of Canada guidelines and regulations, such as "Ethical Conduct for Research Involving Humans (TCPS 2)"

### 2.5 Masked Pathologist Evaluation

To assess the diagnostic quality of the virtual stains, a masked validation study was conducted with three board-certified pathologists. A total of 20 tissue images were selected, consisting of 10 chemically stained images and their corresponding virtual stains from the same tissue samples, ensuring direct one-to-one comparisons. Each image was presented at both low-magnification (overview) and high-magnification (cellular detail). The survey images included all stain types used in this study and incorporated both healthy and disease cases across multiple tissue types. The image order was randomized, rotated, and flipped to minimize recognition of paired samples, and pathologists were masked to image origin.

Pathologists were asked to provide a score for the diagnostic quality of each image on a three-point scale (1 = poor, 2 = good, 3 = excellent). They were also asked to identify whether the image was chemically stained or not, with the options "yes," "no," or "uncertain." Responses were collected independently for all participants.

## 3. Results and Discussion

### 3.1 Expanded Contrast and Biomolecular Specificity with Addition of 355 nm Excitation

The dual-excitation PARS microscope captures label-free contrast through complementary radiative and non-radiative mechanisms at two ultraviolet excitation wavelengths. To illustrate the distinct features captured with each wavelength, Figure 4 overviews images from human skin, human kidney, and mouse kidney samples, and demonstrates the value of combining these contrasts in a single visualization. Across all samples, the 266 nm and 355 nm pulse energies were kept low and measured at 163pJ and 808pJ, respectively, generate strong signal without tissue damage visible in any of the stained samples. Similar to previous work [38], the system achieves a spatial resolution in the 500 nm range, comparable to that of a 40× brightfield scanner [56], allowing detailed subcellular visualizations.

Figure 4(a-b) presents representative images acquired from human skin tissue diagnosed with nodular melanoma. Figure 4a shows a breakdown of the contrast channels obtained at 266 nm (left) and 355 nm (right), with each wavelength producing both non-radiative and radiative signals. This yields four distinct contrast channels per scan, each reflecting different structural and biochemical properties within the tissue. Importantly, because the 266 nm and 355 nm excitations are interlaced within a single scan trajectory, these four contrasts are captured essentially simultaneously, without compromising imaging speed. The 266 nm

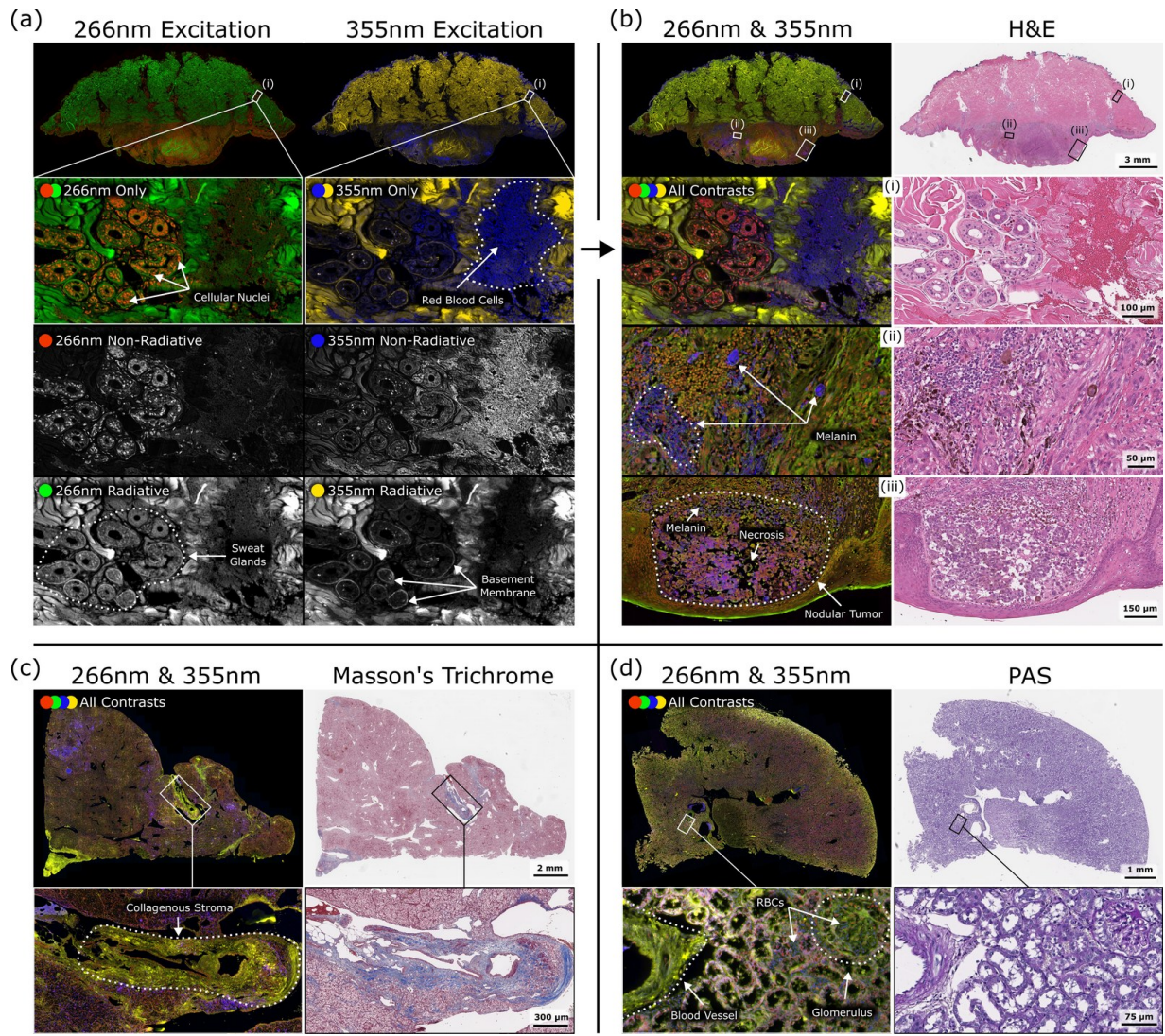

**Figure 4:** Summary of label-free contrasts captured by 266 nm and 355 nm excitation in the PARS system, showing strong correspondence with H&E and specialized stains. (a) PARS images of human skin malignancy (nodular melanoma) showing separate non-radiative (NR) and radiative (R) contrasts at 266 nm and 355 nm. The 266 nm NR highlights nuclei, while 355 nm NR emphasizes RBCs and surrounding tissue detail. Radiative signals from both wavelengths outline connective tissue structures (e.g., sweat glands), with 355 nm better emphasizing the basement membranes. (b) Composite PARS image and corresponding H&E-stained section from the same sample. (b-ii) Melanoma with strong 355 nm NR sensitivity to melanin (blue); (b-iii) Tumor with melanin, necrosis, and vertical growth against the epidermis. (c) PARS image of human kidney with ccRCC, compared to Masson's trichrome. The yellow 355 nm R signal corresponds well to the blue collagen-rich stroma in the stain. (d) PARS image of healthy mouse kidney and PAS stain. The PAS stain highlights carbohydrate-rich structures in a magenta color. Similar features are visible between the PARS and PAS image including nuclei (red, 266 nm NR), RBCs (blue, 355 nm NR), reticular fibres in the blood vessel, and the basement membranes of the glomerulus and surrounding tubules (yellow/green, R signals).

excitation provides strong nuclear contrast through non-radiative absorption of nucleic acids, and radiative signals from connective tissues and fluorophores in the ECM, providing clear visualization of surrounding tissue structures such as sweat glands.

In contrast, 355 nm excitation reveals biologically meaningful tissue features that are less prominent or absent under 266 nm. The non-radiative contrast strongly highlights RBCs through hemoglobin absorption [10] and also provides signal in surrounding tissue structures not as well emphasized with 266 nm non-radiative. The radiative emissions at 355 nm similarly highlights many fluorophores and structures of the ECM (e.g., collagen and elastin), but with different spectral weightings than 266 nm radiative. For example, basement membranes of sweat glands are more pronounced in the 355 nm radiative channel, reflecting differences in absorption and emission efficiencies.

Figure 4b shows the composite PARS image from all four channels, alongside the corresponding H&E-stained section from the same malignant human skin sample. In Figure 4(b-ii), a region of melanoma was selected to demonstrate the strong 355 nm non-radiative sensitivity to melanin. The 355 nm non-radiative signal (blue), clearly outlines heavily pigmented tumour cells, corresponding to dark melanin regions in the H&E image, highlighting the diagnostic value of 355 nm excitation.

Figure 4(b-iii) shows a nodular tumour with characteristic morphology, including nuclear pleomorphism, necrosis, and extensive melanin deposition. The epidermis is clearly delineated, and the tumour mass appears to press against it, consistent with the vertical growth patterns typical of nodular melanoma [57]. The combined 266 nm and 355 nm contrasts align closely with the H&E reference but 355 nm enhances visibility of tumour architecture and melanin accumulation.

Figure 4c shows a human kidney sample with ccRCC and compares the composite PARS image to the corresponding Masson's trichrome-stained section. Masson's trichrome selectively stains collagen-rich regions in blue, helping identify areas of fibrosis and stromal remodeling [4]. These same collagen rich regions are clearly visible in the radiative channels of the PARS image, particularly in the 355 nm autofluorescence response (yellow). While collagen exhibits autofluorescence under both excitation wavelengths, the 355 nm response is more prominent, while the 266 nm radiative signal appears more uniform across the tissue.

Figure 4d shows a healthy mouse kidney sample imaged with PARS and compared to the corresponding PAS-stained section. The PAS stain highlights carbohydrate-rich structures, such as polysaccharides (e.g., glycogen), glycoproteins, and mucosubstances. It is commonly used in kidney pathology to delineate the basement membranes of glomerular capillary loops and renal tubules, which appear magenta [58]. In the PARS image, both radiative channels show the tubule structures, and the capillary loops of the glomerulus correspond to the pink-stained regions in the PAS image. Nuclei in the glomerulus are shown in red (266 nm, non-radiative), and RBCs in the capillaries are visible in blue (355 nm non-radiative). A blood vessel on the left is outlined in yellow and green from both radiative signals and aligns well with the pink reticular fibers outlined in the PAS image.

Overall, the addition of 355 nm excitation expands the structural and biomolecular contrast available in the PARS system, complementing the contrast provided by 266 nm. In many cases, the two wavelengths highlight common tissue features, doing so with different intensity distributions, offering an alternate visualization of shared structures. Together, the combined contrasts correspond well with the patterns seen in both routine H&E and more specialized stains, suggesting their potential utility as input for label-free multi-stain virtual histology.

### 3.2 High-Fidelity Virtual Staining of Diverse Histochemical Stains

The complementary radiative and non-radiative contrasts captured by the 266 nm and 355 nm excitation wavelengths provide diverse and diagnostically relevant structural and biochemical information within the tissue, forming a strong input foundation for virtual staining. Building on the previous section, the following demonstrates the system's ability to replicate a range of histochemical stains across both healthy and diseased samples, using the RegGAN-based staining framework described in Section 2.3 and the full dual-excitation input.

In evaluating the virtual staining performance of the models, three key aspects of the experimental design should be highlighted. First, all testing results presented in this work were generated from entirely unseen WSIs that were held out during model training and validation. This approach tests the model's generalization performance on new tissue sections, mirroring a real-world clinical use case for such a system, and avoiding potential biases from having seen training pairs from the same sample.

Second, because PARS imaging is label-free and operates at low pulse energies that do not visibly damage the tissue, the same slides were chemically stained after imaging. This was core to the training process but is also important for evaluation because it allows direct one-to-one comparisons with the exact chemical counterpart, which is not always feasible in virtual staining studies.

Lastly, histochemically stained WSIs exhibit a wide range of complex tissue structures across many scales, from global patterns to fine cellular details and layouts. It can be challenging to convey staining performance across these scales through isolated crops alone. For this reason, in addition to showing representative areas, WSI virtual and chemical pairs have been made publicly available at the BioImage Archive (see Data Availability). Slides are standard OME-TIFF format and can be viewed and compared with standard pathology viewing software such as QuPath

Figure 5 presents representative virtual staining results across five stain-tissue combinations. Each example includes the label-free PARS input (left), virtual stain (middle), and chemical reference (right), with a zoomed-out whole slide view and high-resolution crop to demonstrate performance across scales. The figure includes human kidney and skin tissues, primarily from disease cases such as ccRCC, melanoma, and fungal infection, along with one mouse kidney example. Standard H&E staining is shown first, followed by three specialized stains (PAS, Masson's trichrome, and JMS) to demonstrate more specific structural and biochemical features. In addition to the disease-focused cases in Figure 5, Supplementary Figure 1 shows results in healthy mouse tissues, demonstrating model performance and PARS contrasts on additional organs, including brain, gastrointestinal (GI) tract, and kidney.

The first two rows of Figure 5 present H&E-stained sections of human kidney and skin. In both cases, the virtual outputs closely resemble their chemical counterpart, properly staining nuclear structures in purple and cytoplastic content and surrounding connective tissue pink. In the malignant kidney tissue (Figure 5a), the virtual stain accurately reproduces the ccRCC tumor architecture, including nests of cells with distinct cell borders and abundant clear cytoplasm (high power area) [58]. The presence of necrosis is clearly visible in the PARS input and is accurately replicated in the virtual stain. This finding is associated with disease aggressiveness. In addition, the thin fibrous septa that separate the tumor nests are clearly visualized in the virtual stain and the fine vascular structures and RBCs within the capillary lumens are well rendered. In the human skin tissue with nodular melanoma (Figure 5b), the virtual stain accurately captures the melanin-laden tumor cells and architecture of the tumor nodule [57]. Within the tumor, the nuclei display marked pleomorphism, and the model is able to reproduce the small nucleoli inside these cells (higher power area). The distribution and localization of melanin within the tumor closely matches the chemical stain, aided by the 355 nm non-radiative absorption contrast input.

The third row, Figure 5c, presents a case of dermatophytosis, a superficial fungal infection of the skin commonly diagnosed using the PAS stain, which highlights polysaccharides in fungal cell walls (e.g., chitin) [59]. In the virtual stain, PAS-positive fungal hyphae are clearly visible as thin magenta-colored filaments clustered along the stratum corneum. The distribution and localization of these fungal organisms accurately match the chemical reference, as do their color and orientation. The virtual output also faithfully reproduces the structure and organization of the surrounding epidermis layers. Notably, the fungal elements are discernible in the label-free PARS input, providing valuable contrast that supports accurate virtual staining of these diagnostically important structures.

In the ccRCC kidney stained with Masson's trichrome (Figure 5d), the virtual output closely matches the chemical reference and enables accurate assessment of the fibrosis surrounding the invasive tumor. The intricate blue and red staining in the collagen-rich area at the bottom right is accurately translated in detail in the virtual stain. In contrast, the tumor growth in the upper left has minimal collagen content, and the virtual stain also accurately reflects this, providing clear visualization of the tumor-stroma interface. The higher power region further highlights the model's sensitivity to subtle collagen cues within the tumor nests, which could aid the evaluation of tumor-driven stromal remodelling.

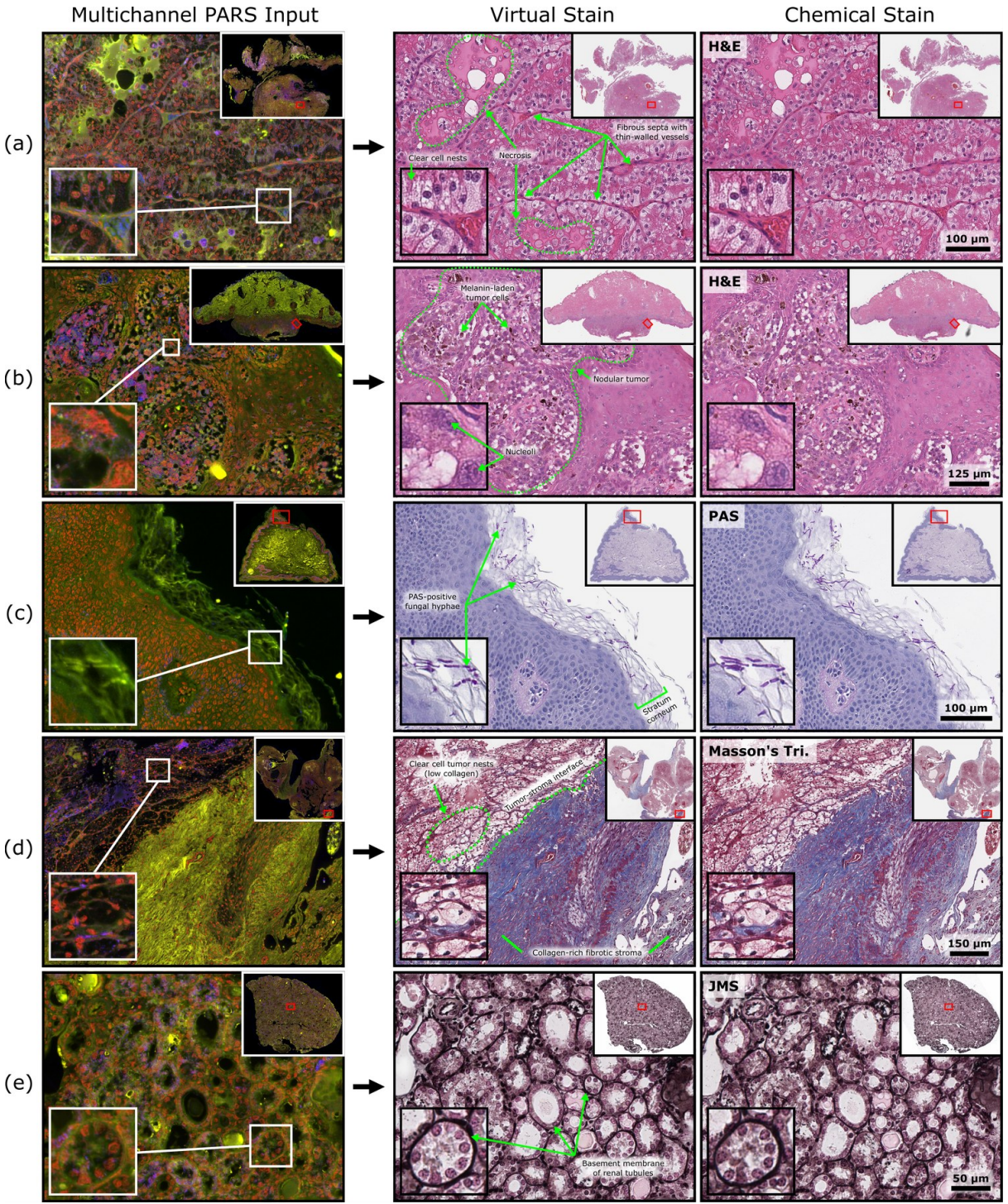

**Figure 5:** Virtual staining results using the dual excitation PARS input and the RegGAN framework across healthy and diseased tissue cases. Examples span multiple tissue types and include both routine H&E and specialized histochemical stains. For each row: (left) label-free PARS input image, (middle) virtual stain, and (right) chemically stained reference. (a) H&E stain of human kidney with ccRCC, showing tumor nests with distinct cell borders and clear cytoplasm, well-defined fibrous septa with a rich capillary network, and regions of necrosis. (b) H&E-stained human skin with nodular melanoma, capturing melanin-laden tumor cells and nuclear pleomorphism within the tumor nodule. (c) PAS stain of human skin with fungal infection, revealing PAS-positive fungal hyphae in the stratum corneum. (d) Masson's trichrome stain of ccRCC kidney, displaying the collagen-rich stroma (bottom right), low-collagen tumor nests (upper left), and a clear tumor-stroma interface. (e) JMS stain of healthy mouse kidney, showing silver-positive (black) tubular basement membranes and accurate coloring of the Nuclear Fast Red counterstain (pink).

The last row, Figure 5e, shows the JMS stain applied to healthy mouse kidney tissue. This stain highlights glomerular and tubular basement membranes in black and is invaluable for the assessment of non-neoplastic kidney biopsies [60]. Overall, the virtual output closely matches the chemical reference, sharply staining the basement membranes between tubules black and accurately coloring the pink nuclear and background counterstain (Nuclear Fast Red). The higher magnification area showcases a distal tubule structure with a well-defined black border and clearly resolved epithelial nuclei, matching well with its chemical reference.

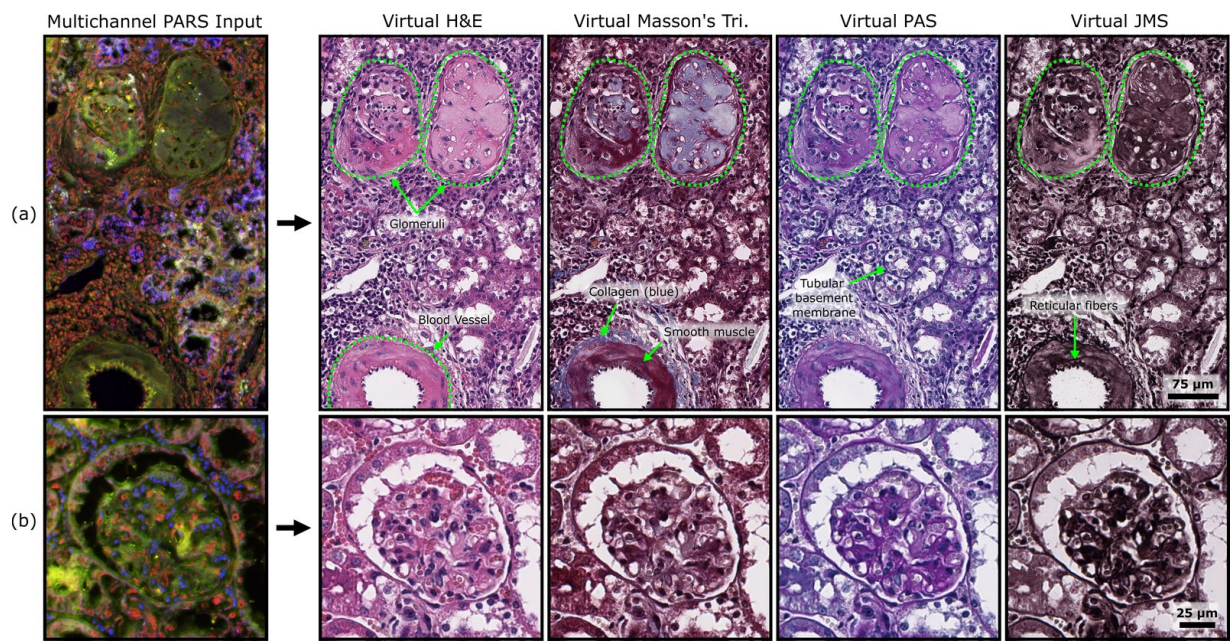

**Figure 6:** Simultaneous virtual H&E, Masson's trichrome, PAS, and JMS staining of mouse kidney tissue from a single PARS input. (a) Region containing two abnormal glomeruli with nodular expansion of mesangial areas, along with a blood vessel and surrounding renal tubules. (b) Normal glomerular architecture from another mouse kidney section, shown in the same four virtual stains for comparison.

To further highlight the utility of the virtual staining models, Figure 6 presents simultaneous virtual H&E, Masson's trichrome, PAS, and JMS generated from a single multichannel PARS input in mouse kidney tissue. Figure 6a highlights a region containing both abnormal glomeruli and surrounding normal structures, while Figure 6b shows a normal glomerulus for comparison. The observations in Figure 6a were made in an aging mouse, where spontaneous, age-related glomerular alterations are not uncommon [61, 62].

The two glomeruli in Figure 6a show globally disrupted architecture, with disrupted and compressed capillary networks due to mesangial area expansion by pale-staining, amorphous material. Each of the virtual stains contributes information about the nature of the glomerular alteration. In the virtual H&E, the glomeruli appear pale pink. The virtual Masson's trichrome shows the mesangial nodules to appear pale blue-gray (versus the darker blue color typical of collagen). The virtual PAS and JMS stains display reduced staining of the amorphous material, with PAS appearing pale and JMS a pink-gray color, lacking the strong magenta color of PAS or the sharply defined black staining of JMS seen in the normal glomerulus in Figure 6b. These combined staining insights are most consistent with glomerular amyloidosis [63], a condition involving extracellular protein deposits (i.e., amyloid) within glomeruli that disrupts glomerular filtration barrier and impairs selective permeability.

The region in Figure 6a also includes several well-preserved normal structures. A blood vessel with clearly defined smooth muscle is visible across all stains, with surrounding collagen clearly identifiable in the virtual Masson's trichrome. Reticular fibers can also be seen within the vessel wall in the virtual JMS (and PAS). The surrounding renal tubules are also properly stained and sharply outlined with the virtual PAS and JMS. Supplementary Figure 2 and 3 provide additional examples of simultaneous virtual multi-staining in healthy mouse GI tissue and human ccRCC kidney tissue, where the multiple stains offer complementary views of tissue structure and pathology.

Together, the example demonstrates how each virtual stain contributes distinct structural and biochemical information that, when combined, allows for a more complete interpretation of the glomerular pathology than routine H&E or any single stain alone could provide. For example, work by Haan et al. demonstrated that virtual multi-staining, achieved through stain-to-stain transformations, can improve diagnostic accuracy in renal pathology, supporting the value of combining multiple stain types [64]. In addition to the virtual stains, the label-free PARS input here offers an additional distinct visualization of tissue structures, with a unique contrast mechanism independent from chemical staining. Moreover, because the sample remains unaltered, subsequent staining with other specialized markers can be performed. For example, to definitively confirm amyloid deposition in Figure 6a, the same kidney section could be chemically stained with Congo red and imaged under polarized light microscopy [63].

## 3.3 Pathologist Validation Results

The results from the masked pathologist evaluation further demonstrate the close correspondence between chemical and virtual stains. As detailed in section 2.5, expert pathologists, masked to the image origin, reviewed a set of randomized chemical and virtual image pairs spanning all stain types, tissue types, and disease contexts included in this study.

**Table 2:** Summary of masked pathologist evaluation of chemical and virtual stains.

| Image Origin | Avg. DQ (± Std) | Marked Chemical | Marked Virtual | Uncertain |
|---|---|---|---|---|
| Chemical | 2.600 ± 0.306 | 17 | 1 | 12 |
| Virtual | 2.667 ± 0.314 | 18 | 1 | 11 |

DQ = Diagnostic Quality (1: poor, 2: good, 3: excellent).

A summary of diagnostic quality scores and image original classification is provided in Table 2, with the full responses presented in Supplemental Table 1. Diagnostic quality scores were comparable between the chemical and virtual images, with both image types consistently scored as "good" or "excellent". In terms of image classification, pathologists were unable to reliably determine image origin. The majority of images (both chemical or virtual) were marked as chemical (17 chemical and 18 virtual), while only a single image of each type was marked as virtual. Furthermore, 12 chemical and 11 virtual images were judged as uncertain. One pathologist correctly identified a single virtual image but also misclassified a chemical image as virtual, with their remaining responses marked as chemical or uncertain. Overall, the masked evaluation confirms two key outcomes: (1) diagnostic quality of the virtual stains was preserved across diverse tissues, stains, and disease contexts, and (2) virtual and chemical images could not be reliably separated by experienced pathologists. Together, these results demonstrate that dual-excitation PARS virtual staining preserves diagnostic quality while producing images that closely match their chemical counterparts.

## 3.4 Contribution of Individual Excitation Channels to Virtual Staining Performance

To evaluate the individual and combined contributions of 266 nm and 355 nm excitation contrasts, additional models were trained using only 266 nm input and only 355 nm input. These were compared to the dual excitation models presented in the previous section for all tissue and stain combinations used in this study (Table 1). Relative performance was then quantified using two metrics: (1) Multi-Scale Structural Similarity Index (MS-SSIM) [65], which captures pixel-level structural similarity across scales, with higher values indicating better similarity; and (2) Deep Image Structure and Texture Similarity (DISTS) [66], which captures perceptual similarity based on structural and textural fidelity, with lower values indicating better similarity. The overall aggregated results across all tissue and stain combinations are summarized in Table 3. For completeness, the per-dataset breakdown of these metrics is provided in Supplementary Table 2 and Supplementary Table 3.

**Table 3:** Summary quantitative scores (Avg. ± Std) across all datasets for different excitation wavelength input combinations.

| Model Input | DISTS ($\downarrow$) | MS-SSIM ($\uparrow$) |
| --- | --- | --- |
| Dual Excitation | 0.138 ± 0.027 | 0.904 ± 0.063 |
| 266 nm Only | 0.145 ± 0.027 | 0.879 ± 0.070 |
| 355 nm Only | 0.150 ± 0.027 | 0.862 ± 0.078 |

*Note*: Metrics are computed on 256×256px patches, N=292,664.

In all cases, models trained with both excitation contrasts achieved improved scores (higher for MS-SSIM and lower for DISTS) compared to those trained with either wavelength alone. These results are consistent with the idea that the complementary contrast from both wavelengths provides more biochemical specificity and structural information to the model, supporting improved virtual staining performance. In general, 266 nm models outperformed 355 nm models for most tissue-stain combinations, likely attributed to the essential nuclear contrast present in the 266 nm non-radiative channel. One notable exception was observed in the skin melanoma sample, where the 355 nm model achieved a better DISTS score, reflecting improved replication of melanoma texture from the 355 nm non-radiative channel.

While these metrics help assess overall staining quality, further analysis was done to assess how the addition of 266 nm or 355 nm affects the virtual staining of more specific histological features. Figure 7 presents three representative examples, collagen in Masson's trichrome, nuclei in H&E, RBCs, and fungal hyphae in PAS-stained skin, with each demonstrating how dual excitation improves staining of clinically relevant features.

Figure 7a compares virtual Masson's trichrome staining of ccRCC kidney tissue using models trained with 266 nm only versus dual excitation input. Using QuPath digital pathology software, a pixel classifier was trained to segment collagen-positive pixels. The collagen proportionate area (CPA), defined as the fraction of tissue area classified as collagen, was computed across 256×256 px patches in regions with at least 5% collagen. While CPA is primarily used in liver biopsies to assess fibrosis and clinical outcomes [67], here it serves as a quantitative proxy for evaluating collagen staining accuracy. Three example regions are shown in Figure 7a, in order of increasing CPA values. In each example, the 266 nm model underestimates collagen presence, mis-staining collagen areas in red, leading to significantly reduced CPA values. By adding the 355 nm channels, the model more closely matches the reference stain, as reflected in the segmentation maps (bottom left), which show a similar distribution of collagen. These qualitative findings are further supported in the violin plots below, which show CPA distributions across N = 4,205 patches. The dual excitation model achieves an almost symmetric distribution (median 0.57 ± 0.25), closely matching the ground truth (0.59 ± 0.25). In contrast, the 266 nm model exhibits a skewed distribution (median 0.40 ± 0.25), reflecting an overall underestimation of collagen. The improvement likely results from the addition of the 355 nm radiative channel, which provides complementary autofluorescence contrast to 266 nm. While collagen fluoresces under both excitations, many surrounding structures do as well. Access to both radiative channels allow the model to better stain collagen based on learned differences in emission intensities across tissue structures. Surrounding tissue structures also exhibit non-radiative contrast at 355 nm, which may further aid in distinguishing collagen.

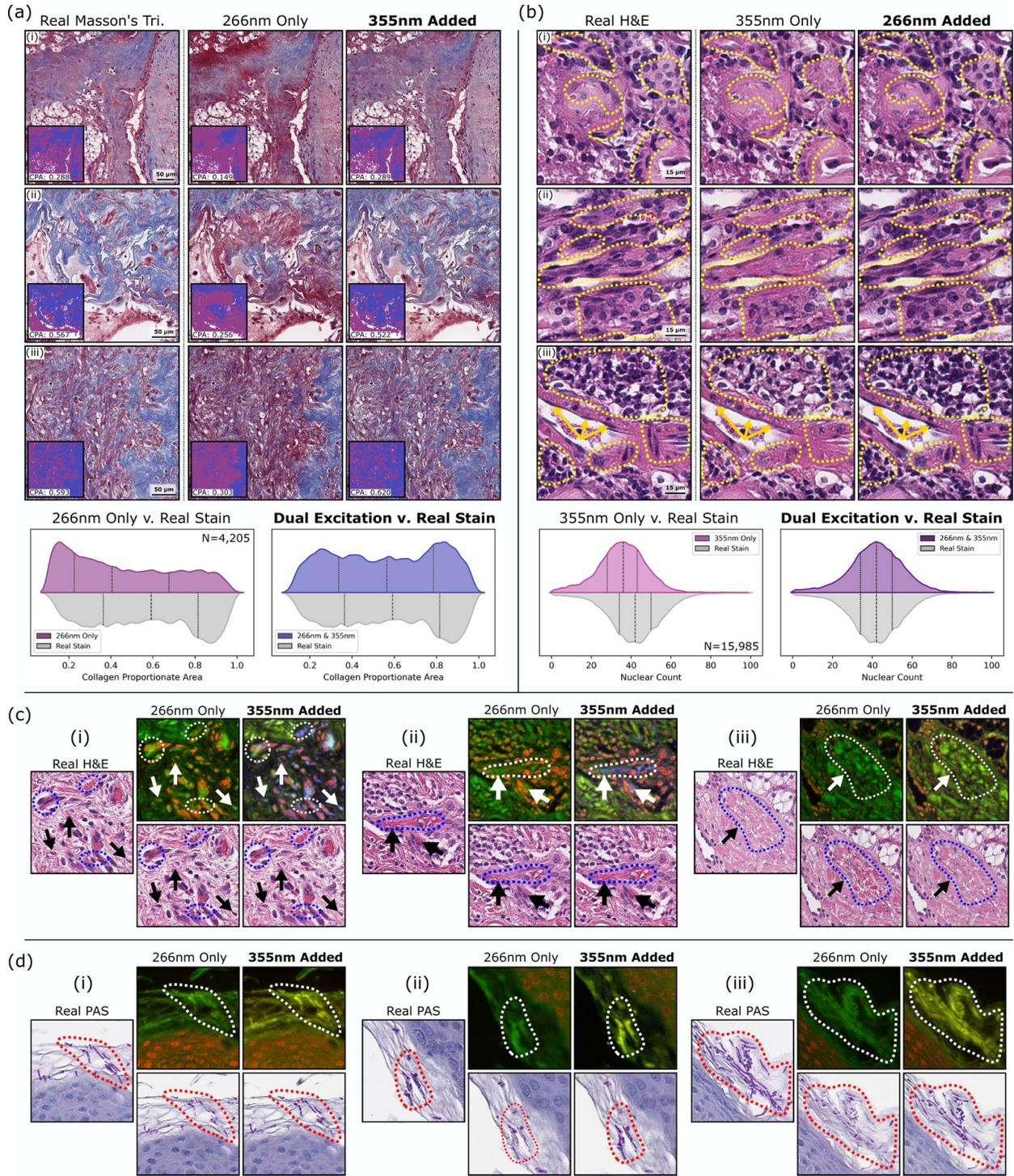

**Figure 7:** Dual-excitation models improve staining of histological features across multiple stains. (a) Masson's trichrome, ccRCC kidney: Each region shows a collagen segmentation map (bottom left) and Collagen Proportionate Area (CPA). Violin plots of CPA distribution (N=4,205, 256px patches) show the dual excitation model (median 0.57) closely matches the reference (0.59), versus the 266 nm only model (0.40). (b) H&E, mouse kidney: Dashed outlines and arrows highlight missing nuclei in the 355 nm only output. Adding 266 nm contrast improves nuclear staining. Violin plots of nuclear count distribution (N=17,821, 256px patches) show close agreement between the dual excitation model (median 42.00) and the reference (41.57), compared to the 355 nm-only model (34.95). (c) Improvement of RBC staining with addition of 355 nm excitation. In (c-i) and (c-ii), RBCs are absent or miscolored in the 266 nm only model; in (c-iii), false RBC-like structures appear where none exist in the reference. (d) PAS, skin tissue: Dual excitation improves fungal hyphae staining. Adding 355 nm results in more complete hyphae staining, whereas the 266 nm only model shows incomplete or missed structures.

Figure 7b demonstrates the benefits of including 266 nm input for nuclear staining. Examples are shown in H&E mouse kidney and compare the 355 nm only model to the dual excitation output. In each case, the 355 nm model struggles to accurately reconstruct nuclei, and several nuclear structures are missed entirely and replaced by pink eosin staining. Although 355 nm lacks nuclear contrast, the model does successfully infer some nuclear structures from the surrounding tissue, but this is not always reliable. In contrast, by adding 266 nm excitation, the model consistently identifies and correctly stains nuclei. The quantitative nuclear counts reinforce this difference. Using StarDist [68] for segmentation, nuclear counts were computed for N=17,821 patches. The dual excitation model closely matches the chemical H&E distribution (median 42.00 ± 14.50 vs. real 41.57 ± 14.23), while the 355 nm model shows a similar distribution shape but with a shifted center, reflecting the overall undercounting of nuclei (34.95 ± 12.54).

Figure 7c shows three examples of how the addition of 355 nm channels improves staining of RBCs. While the 266 nm model can often identify RBCs using structural cues, it lacks hemoglobin-specific contrast which can lead to errors. In Figure 7 examples (c-i) and (c-ii), RBCs are either missed or incorrectly stained purple as nuclei. While in Figure 7(c-iii), the 266 nm model hallucinates red RBC-looking structures where none exist. In contrast, the 355 nm non-radiative channel (shown in blue) captures hemoglobin absorption, providing both positive and negative cues that help the model more accurately identify RBCs and avoid such false positives.

Lastly, Figure 7d shows three examples of PAS staining of fungal hyphae in skin tissue. The 266 nm model identifies some fungal structures but often underrepresents or only partially stains them. With both excitation contrasts, the model more accurately stains the fungal elements, aided by the 355 nm radiative channel (shown in yellow), which enhances their visibility and allows for more complete staining. In each case, the dual excitation output more closely matches the reference PAS stain. While some deviations remain and the staining is not perfect, the addition of 355 nm contrast leads to a noticeably improved staining of fungal hyphae.

### 3.5 Comparative Performance of Virtual Staining Models

The RegGAN model was chosen for virtual staining because it enables supervised training while remaining robust to imperfect alignments by integrating a registration network during training. This makes it well-suited for cross-modality tasks where precise pixel-level alignment is difficult to achieve. To evaluate its effectiveness, RegGAN was benchmarked against two commonly used approaches: Pix2Pix (paired) and CycleGAN (unpaired). Figure 8 presents qualitative and quantitative comparisons across the models, using a subset of examples that highlight each stain applied to a relevant tissue type or disease case featured in this study. Figure 8(a-e) show: (a) H&E-stained human kidney with ccRCC (N=60,531), (b) H&E-stained human skin with nodular melanoma (N=21,739), (c) PAS-stained human skin with fungal infection (N=11,406), (d) Masson's trichrome-stained kidney with ccRCC (N=66,672), and (e) JMS-stained healthy mouse kidney (N=16,473). Each row in the figure includes the 266 nm and 355 nm PARS input, the corresponding chemical stain, and outputs from the three models. In addition, violin plots showing the relative performance of each model are provided for both the DISTS (perceptual similarity) and MS-SSIM (structural similarity) metrics. The overall aggregated quantitative scores for RegGAN, Pix2Pix, and CycleGAN across all tissue–stain combinations are summarized in Table 4. Detailed per-dataset results are reported in Supplementary Tables 4 and 5.

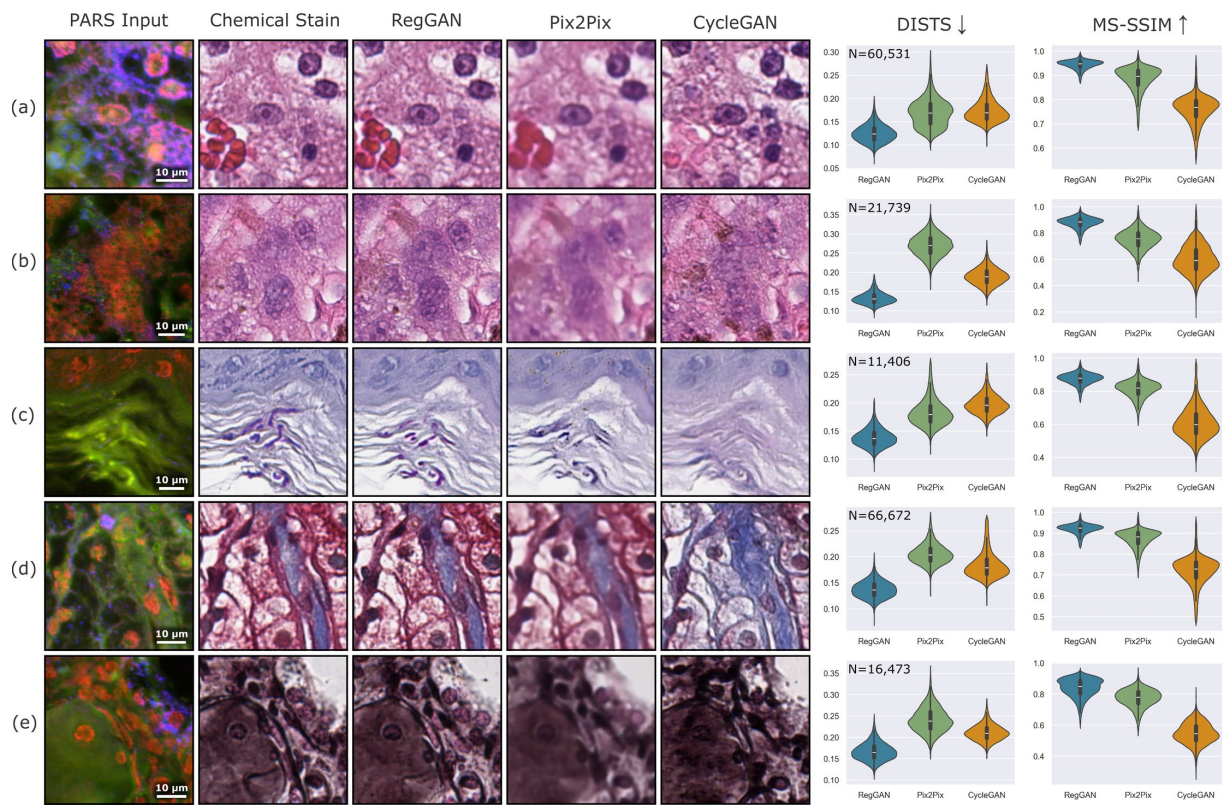

**Figure 8:** Visual and quantitative comparison of virtual staining performance for RegGAN, Pix2Pix, and CycleGAN models. Each example shows the PARS input image, ground truth chemical stain, and outputs from all three models, along with distributions of quantitative metrics (DISTS and MS-SSIM). The comparison covers all stains used in this study and includes representative disease cases where possible: (a) H&E-stained human kidney with ccRCC, (b) H&E-stained human skin with nodular melanoma, (c) PAS-stained human skin with fungal infection, (d) Masson's trichrome-stained human kidney with ccRCC, and (e) JMS-stained healthy mouse kidney.

**Table 4:** Summary quantitative scores (Avg. ± Std) for RegGAN, Pix2Pix, and CycleGAN models across all datasets.

| Model | DISTS (↓) | MS-SSIM (↑) |
|---|---|---|
| RegGAN | 0.138 ± 0.027 | 0.904 ± 0.063 |
| Pix2Pix | 0.194 ± 0.047 | 0.843 ± 0.082 |
| CycleGAN | 0.193 ± 0.035 | 0.663 ± 0.134 |

*Note*: Metrics are computed on 256×256px patches, N=292,664.
20

The Pix2Pix model, a strictly supervised model trained on aligned pairs, accurately captures the overall stain color and tissue structure. However, in each example in Figure 8, it consistently generates overly smoothed and blurry outputs, failing to reproduce the correct textures and structural detail. While it may be sufficient for low magnifications, the blurring limits diagnostic utility at higher magnifications, where features like mitotic figures, nuclear atypia, and subtle architectural cues are important. These limitations are evident in the loss of nuclear detail in Figure 8b, poorly defined glomerulus basement membranes in the JMS-stained kidney (Figure 8d), and incomplete PAS fungal staining in Figure 8c. Despite careful registration using tools like Warpy, Pix2Pix remains sensitive to slight misalignments and differences between imaging modalities, which are inevitable in real-world datasets. As well, its performance would likely degrade significantly with noisier or less curated inputs.

The CycleGAN model, trained in an unpaired fashion, produces sharper images and retains textural detail more effectively. However, its less constrained training approach limits its ability to learn more complex contrast mappings. For example, RBCs are poorly reproduced in the H&E kidney (Figure 8a); the blue collagen stain is exaggerated and poorly localized in the trichrome example (Figure 8d); fungal elements are entirely missed in the PAS skin (Figure 8c); and melanin and nuclei are poorly rendered in the H&E melanoma case (Figure 8b). In the JMS-stained kidney (Figure 8e), CycleGAN roughly applies the correct stain color and texture but fails to accurately stain the real basement membrane structures.

In contrast, by incorporating a registration network into the training process, RegGAN benefits from paired training feedback while being robust to data misalignments. It preserves high-resolution texture and structure better than Pix2Pix, particularly evident in the accurate rendering of nuclear detail (Figure 8b) and basement membranes (Figure 8e), while also achieving more faithful stain translation than CycleGAN. It successfully captures collagen in the trichrome example (Figure 8d), the PAS-stained fungal elements (Figure 8c), and melanin in the skin example (Figure 8b), producing the most consistent alignment with the chemical reference across all examples.

These qualitative observations are supported by the quantitative metrics. RegGAN achieves the best performance across all stain–tissue combinations under the DISTS metric, which captures perceptual similarity based on structural and textural fidelity (lower values reflect better similarity). CycleGAN generally achieves better DISTS scores than Pix2Pix, particularly in Figure 8 (b), (d), and (e), where Pix2Pix exhibits more pronounced blurring and textural loss. However, CycleGAN underperforms in specialized staining tasks such as PAS fungal staining (Figure 8c), where it fails to capture key features that Pix2Pix more accurately reproduces.

When evaluated with MS-SSIM, which captures pixel-level structural similarity across scales (higher values reflect better similarity), RegGAN again outperforms both models. CycleGAN performs the worst in all MS-SSIM examples, reflecting its failure to accurately map certain structural features and stain distributions, despite its sharper overall appearance and textural similarity.

RegGAN demonstrated the strongest performance in virtual staining; however, the results highlight the broader utility of incorporating registration directly into a training pipeline. Its ability to handle small but unavoidable misalignments reduces the time burden of precise data curation. It also helps mitigate challenges in cross-modality learning tasks, particularly those involving mismatched resolutions or differences in acquisition, such as optical sectioning, illumination, or other imaging conditions. By addressing these discrepancies as part of the training and optimization process, registration-aware approaches enable more scalable and robust use of imperfectly paired data. Beyond virtual staining, this strategy may benefit a range of cross-modality learning tasks in biomedical imaging where alignment is a limiting factor.

## 4. Conclusion

This study presents a dual-excitation PARS microscopy system for label-free virtual multi-stain histology, combining 266 nm and 355 nm UV excitations to capture complementary radiative and non-radiative con-trasts. The addition of 355 nm expands the range of histological features that can be distinguished, including RBCs and melanin-containing elements, and enhances visualization of stromal architecture including collagen and elastin, while 266 nm excitation provides important nuclear contrast and surrounding connective tissue structures. Using the RegGAN image-translation framework, this study presents the first PARS virtual staining of specialized stains such as Masson's trichrome, PAS, and JMS, alongside H&E, across healthy and pathological human and murine tissues, including kidney, skin, brain, and gastrointestinal tract. Evaluation on unseen WSIs showed close correspondence between virtual and chemical stains, with improved similarity metrics, perceptual metrics, and more accurate reconstruction of histology-specific features when both excitation wavelengths were used. A masked pathologist evaluation further demonstrated that virtual stains preserved diagnostic quality comparable to their chemical counterparts, and pathologists were unable to reliably determine image origin. Together, these results establish dual-excitation PARS virtual staining as a non-destructive addition to the histopathology workflow, enabling pre-stain histology imaging that preserves tissue for conventional histology and downstream molecular analyses, while expanding diagnostic insight.

**Data Availability**

Representative WSI pairs have been uploaded to the BioImage Archive. For each tissue–stain combination included in this study, corresponding real and virtual WSIs are publicly available under accession number *S-BIAD2232*, accessible at https://doi.org/10.6019/S-BIAD2232. All uploaded WSIs were entirely held out from the training and validation sets. Data are provided in OME-TIFF format and can be viewed with standard pathology software such as QuPath.

**Funding**

This research was funded by: Natural Sciences and Engineering Research Council of Canada (DGECR-2019-00143, RGPIN2019-06134, DH-2023-00371); Canada Foundation for Innovation (JELF #38000); Mitacs Accelerate (IT13594); University of Waterloo Startup funds; Centre for Bioengineering and Biotechnology (CBB Seed fund); illumiSonics Inc (SRA #083181); New Frontiers in Research Fund – Exploration (NFRFE-2019-01012); The Canadian Institutes of Health Research (CIHR PJT 185984), (PJT-195962).

**Acknowledgements**

The authors thank Dr. Marie Abi Daoud (Alberta Precision Laboratories, Calgary, Canada) for providing human skin tissue samples with fungal infection, and Dr. Ally Khan Somani (Indiana University School of Medicine) for providing human skin tissue samples with nodular melanoma. Additionally, the authors would like to acknowledge Hager Gaouda and Sabeena Giri for their assistance in staining some of the tissue samples used in this study.

The authors also thank Dr. Hector Li-Chang, Dr. Will Chen and Dr. Bibi Naghibi Torbati for participating in the stain quality assessment survey. In addition, the authors thank Dr. Hector Li-Chang and Dr. Jamie Lee who helped review and provide input on the histology annotations and descriptions used in this work.

Biological materials (human kidney samples) were provided by the Ontario Tumour Bank, which is supported by the Ontario Institute for Cancer Research through funding provided by the Government of Ontario. The views expressed in this publication are the views of the authors and do not necessarily reflect those of the Government of Ontario.

**Competing Interests**

Authors James E.D. Tweel, Benjamin R. Ecclestone, James A. Tummon Simmons, and Parsin Haji Reza all have financial interests in IllumiSonics, which has provided funding to the PhotoMedicine Labs.

# 5. Supplemental Information

**Supplementary Table 1:** Results of masked pathologist evaluation showing responses for diagnostic quality (DQ) and image origin (IO).

| Type – Stain – Sample | Path. 1 | | Path. 2 | | Path. 3 | | Avg. DQ |
|---|---|---|---|---|---|---|---|
| | DQ | IO | DQ | IO | DQ | IO | |
| 1. C – H&E – Kidney ccRCC | 3 | Y | 2 | U | 3 | U | 2.67 |
| 1. V – H&E – Kidney ccRCC | 3 | Y | 2 | Y | 3 | U | 2.67 |
| 2. C – MT – Kidney ccRCC | 2 | Y | 2 | Y | 3 | U | 2.33 |
| 2. V – MT – Kidney ccRCC | 2 | Y | 2 | U | 3 | U | 2.33 |
| 3. C – PAS – Mouse Kidney | 3 | Y | 1 | N | 3 | U | 2.33 |
| 3. V – PAS – Mouse Kidney | 3 | Y | 2 | Y | 3 | U | 2.67 |
| 4. C – JMS – Mouse Kidney | 3 | Y | 1 | Y | 2 | U | 2.00 |
| 4. V – JMS – Mouse Kidney | 2 | Y | 2 | Y | 2 | U | 2.00 |
| 5. C – H&E – Skin Melanoma | 3 | Y | 3 | Y | 3 | U | 3.00 |
| 5. V – H&E – Skin Melanoma | 3 | Y | 3 | Y | 3 | U | 3.00 |
| 6. C – PAS – Mouse GI | 3 | Y | 2 | U | 3 | U | 2.67 |
| 6. V – PAS – Mouse GI | 3 | Y | 2 | N | 3 | U | 2.67 |
| 7. C – MT – Kidney ccRCC | 3 | Y | 2 | Y | 3 | U | 2.67 |
| 7. V – MT – Kidney ccRCC | 3 | Y | 3 | Y | 3 | U | 3.00 |
| 8. C – PAS – Skin Fungal | 3 | Y | 3 | Y | 3 | U | 3.00 |
| 8. V – PAS – Skin Fungal | 3 | Y | 3 | Y | 3 | U | 3.00 |
| 9. C – H&E – Kidney ccRCC | 3 | Y | 2 | Y | 3 | U | 2.67 |
| 9. V – H&E – Kidney ccRCC | 3 | Y | 2 | Y | 3 | U | 2.67 |
| 10. C – MT – Kidney ccRCC | 3 | Y | 2 | Y | 3 | U | 2.67 |
| 10. V – MT – Kidney ccRCC | 3 | Y | 2 | Y | 3 | U | 2.67 |

**DQ** = Diagnostic Quality (1: poor, 2: good, 3: excellent).
**IO** = Image Origin (Y: Yes, N: No, U: Uncertain).
**C** = Chemical stain. **V** = Virtual stain. **MT** = Masson's Trichrome;

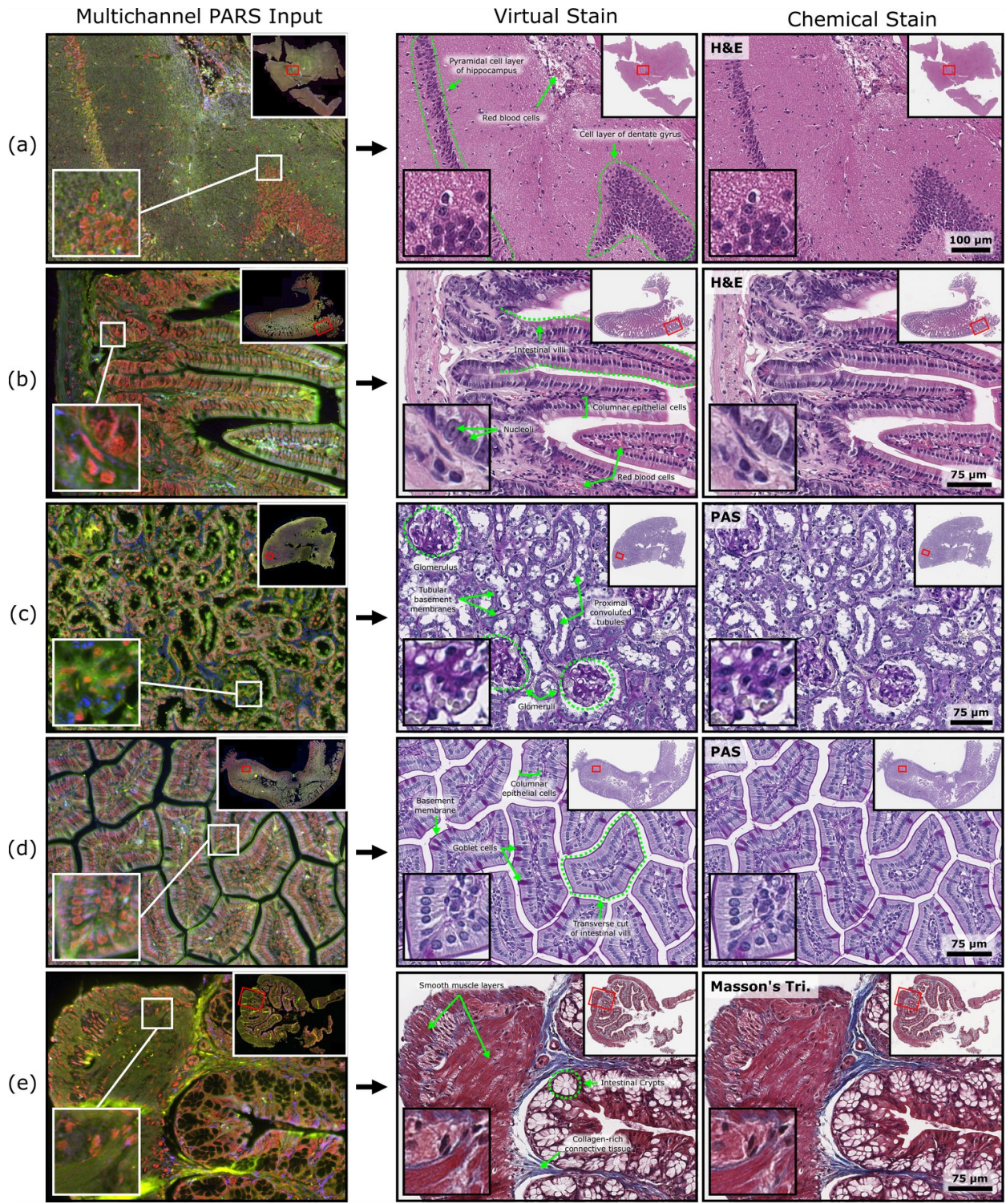

**Supplementary Figure 1:** Virtual staining results using the dual excitation PARS input and the RegGAN framework across healthy mouse tissues. Examples span multiple organ systems and histochemical stains, with labeled annotations highlighting key structural features. For each row: (left) label-free PARS input image, (middle) virtual stain, and (right) chemically stained reference. (a) H&E mouse brain, showing the pyramidal cell layer of the hippocampus, the dentate gyrus, and various RBCs. (b) H&E mouse intestine, revealing intestinal villi, columnar epithelial cells, and higher power section showing nucleoli. (c) PAS mouse kidney, showing glomerular basement membranes (high power section) in magenta and comparable renal tubular appearance between chemical and virtual stains. (d) PAS mouse intestine, displaying transverse sections of intestinal villi and goblet cells. (e) Masson's trichrome stain of mouse intestine, demonstrating smooth muscle layers, collagen-rich connective tissue (blue), and intestinal crypts.

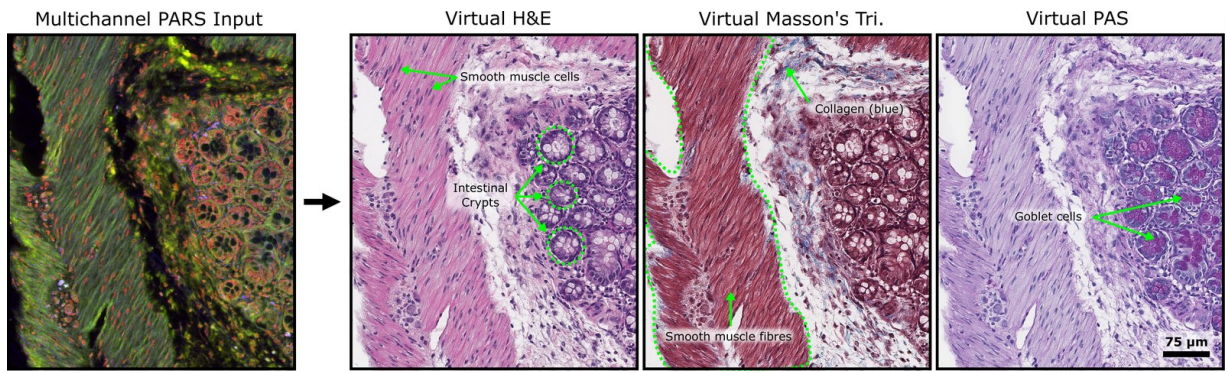

**Supplementary Figure 2:** Virtual histochemical multi-staining of mouse intestine tissue from a single multichannel PARS input. The single PARS image (left) is used to generate three virtual stains: H&E, Masson's trichrome, and PAS. The virtual H&E highlights smooth muscle cells and intestinal crypts. Masson's trichrome shows collagen (blue) and smooth muscle fibers, while PAS reveals magenta-stained goblet cells.

**Supplementary Table 2:** DISTS (↓) scores across datasets for different excitation wavelength input combinations. Each cell reports the Avg. ± Std results.

| Dataset | Test Samples | DISTS (↓) | | |
|---|---|---|---|---|
| | | **Dual Excitation** | **266 nm Only** | **355 nm Only** |
| Kidney ccRCC (H&E) | 60,531 | 0.126 ± 0.023 | 0.136 ± 0.023 | 0.136 ± 0.023 |
| Skin Melanoma (H&E) | 21,739 | 0.135 ± 0.021 | 0.151 ± 0.022 | 0.149 ± 0.022 |
| Skin Fungal Disease (PAS) | 11,406 | 0.142 ± 0.032 | 0.150 ± 0.032 | 0.154 ± 0.033 |
| Kidney ccRCC (Masson's) | 66,672 | 0.139 ± 0.022 | 0.142 ± 0.022 | 0.150 ± 0.022 |
| Mouse Kidney (JMS) | 16,473 | 0.168 ± 0.023 | 0.174 ± 0.023 | 0.174 ± 0.024 |
| Mouse Kidney (H&E) | 17,821 | 0.129 ± 0.020 | 0.136 ± 0.018 | 0.147 ± 0.020 |
| Mouse Kidney (PAS) | 21,594 | 0.128 ± 0.018 | 0.135 ± 0.017 | 0.139 ± 0.017 |
| Mouse Kidney (Masson's) | 18,867 | 0.158 ± 0.032 | 0.163 ± 0.032 | 0.169 ± 0.033 |
| Mouse GI (H&E) | 7,768 | 0.127 ± 0.028 | 0.136 ± 0.026 | 0.149 ± 0.028 |
| Mouse GI (PAS) | 28,345 | 0.145 ± 0.025 | 0.152 ± 0.024 | 0.160 ± 0.025 |
| Mouse GI (Masson's) | 12,996 | 0.148 ± 0.042 | 0.158 ± 0.041 | 0.164 ± 0.042 |
| Mouse Brain (H&E) | 8,452 | 0.130 ± 0.023 | 0.139 ± 0.023 | 0.144 ± 0.024 |

Note: GI, gastrointestinal. Metric are computed on 256×256px patches.

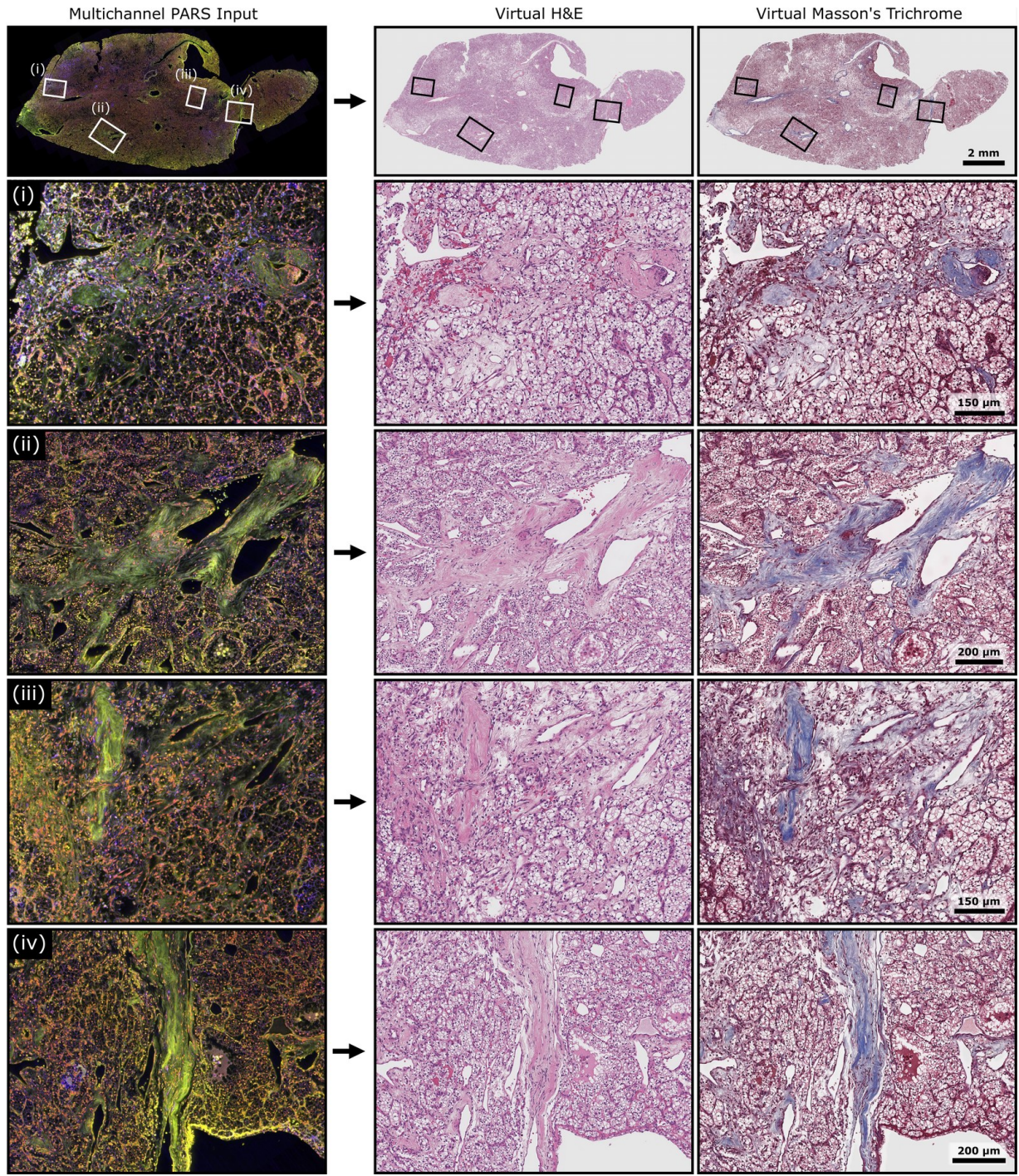

**Supplementary Figure 3:** Virtual histochemical multi-staining of human kidney tissue with ccRCC. A single multichannel PARS image is used to generate two virtual stains: H&E and Masson's trichrome. Whole-slide virtual stains are shown at the top, with four representative regions (i)-(iv) selected for detailed viewing. In the virtual H&E, clear cell tumor nests are visible across the sampled regions, showing characteristic cytoplasmic clearing and nested morphology. The corresponding Masson's trichrome images highlight tumor-driven stromal remodeling, with collagen deposition (blue) surrounding the tumor regions.

**Supplementary Table 3:** MS-SSIM (↑) scores across datasets for different excitation wavelength input combinations. Each cell reports the Avg. ± Std results.

| Dataset | Test Samples | MS-SSIM (↑) | | |
|---|---|---|---|---|
| | | Dual Excitation | 266 nm Only | 355 nm Only |
| Kidney ccRCC (H&E) | 60,531 | 0.939 ± 0.044 | 0.873 ± 0.077 | 0.750 ± 0.079 |
| Skin Melanoma (H&E) | 21,739 | 0.870 ± 0.072 | 0.744 ± 0.088 | 0.594 ± 0.112 |
| Skin Fungal Disease (PAS) | 11,406 | 0.869 ± 0.066 | 0.808 ± 0.102 | 0.613 ± 0.109 |
| Kidney ccRCC (Masson's) | 66,672 | 0.913 ± 0.049 | 0.864 ± 0.062 | 0.707 ± 0.090 |
| Mouse Kidney (JMS) | 16,473 | 0.836 ± 0.069 | 0.771 ± 0.065 | 0.547 ± 0.071 |
| Mouse Kidney (H&E) | 17,821 | 0.926 ± 0.025 | 0.856 ± 0.042 | 0.656 ± 0.081 |
| Mouse Kidney (PAS) | 21,594 | 0.914 ± 0.039 | 0.814 ± 0.046 | 0.667 ± 0.070 |
| Mouse Kidney (Masson's) | 18,867 | 0.896 ± 0.061 | 0.875 ± 0.067 | 0.743 ± 0.097 |
| Mouse GI (H&E) | 7,768 | 0.887 ± 0.039 | 0.871 ± 0.055 | 0.732 ± 0.098 |
| Mouse GI (PAS) | 28,345 | 0.864 ± 0.075 | 0.806 ± 0.090 | 0.476 ± 0.163 |
| Mouse GI (Masson's) | 12,996 | 0.893 ± 0.067 | 0.837 ± 0.085 | 0.641 ± 0.168 |
| Mouse Brain (H&E) | 8,452 | 0.929 ± 0.058 | 0.848 ± 0.089 | 0.592 ± 0.100 |

Note: GI, gastrointestinal. Metric are computed on 256×256px patches.

**Supplementary Table 4:** Quantitative comparison of RegGAN, Pix2Pix, and CycleGAN for each dataset using DISTS (↓). Each cell reports the Avg. ± Std results.

| Dataset | Test Samples | DISTS (↓) | | |
|---|---|---|---|---|
| | | RegGAN | Pix2Pix | CycleGAN |
| Kidney ccRCC (H&E) | 60,531 | 0.126 ± 0.023 | 0.171 ± 0.034 | 0.176 ± 0.031 |
| Skin Melanoma (H&E) | 21,739 | 0.135 ± 0.021 | 0.271 ± 0.030 | 0.191 ± 0.024 |
| Skin Fungal Disease (PAS) | 11,406 | 0.142 ± 0.032 | 0.187 ± 0.068 | 0.199 ± 0.034 |
| Kidney ccRCC (Masson's) | 66,672 | 0.139 ± 0.022 | 0.207 ± 0.024 | 0.190 ± 0.036 |
| Mouse Kidney (JMS) | 16,473 | 0.168 ± 0.023 | 0.242 ± 0.030 | 0.213 ± 0.025 |
| Mouse Kidney (H&E) | 17,821 | 0.129 ± 0.020 | 0.186 ± 0.030 | 0.179 ± 0.019 |
| Mouse Kidney (PAS) | 21,594 | 0.128 ± 0.018 | 0.152 ± 0.023 | 0.188 ± 0.015 |
| Mouse Kidney (Masson's) | 18,867 | 0.158 ± 0.032 | 0.195 ± 0.050 | 0.191 ± 0.031 |
| Mouse GI (H&E) | 7,768 | 0.127 ± 0.028 | 0.175 ± 0.050 | 0.191 ± 0.026 |
| Mouse GI (PAS) | 28,345 | 0.145 ± 0.025 | 0.182 ± 0.033 | 0.222 ± 0.037 |
| Mouse GI (Masson's) | 12,996 | 0.148 ± 0.042 | 0.208 ± 0.068 | 0.236 ± 0.051 |
| Mouse Brain (H&E) | 8,452 | 0.130 ± 0.023 | 0.184 ± 0.031 | 0.191 ± 0.023 |

Note: GI, gastrointestinal. Metric are computed on 256×256px patches.

**Supplementary Table 5:** Quantitative comparison of RegGAN, Pix2Pix, and CycleGAN for each dataset using MS-SSIM (↑). Each cell reports the Avg. ± Std results.

| Dataset | Test Samples | MS-SSIM (↑) | | |
|---|---|---|---|---|
| | | **RegGAN** | **Pix2Pix** | **CycleGAN** |
| Kidney ccRCC (H&E) | 60,531 | 0.939 ± 0.044 | 0.873 ± 0.077 | 0.750 ± 0.079 |
| Skin Melanoma (H&E) | 21,739 | 0.870 ± 0.072 | 0.744 ± 0.088 | 0.594 ± 0.112 |
| Skin Fungal Disease (PAS) | 11,406 | 0.869 ± 0.066 | 0.808 ± 0.102 | 0.613 ± 0.109 |
| Kidney ccRCC (Masson's) | 66,672 | 0.913 ± 0.049 | 0.864 ± 0.062 | 0.707 ± 0.090 |
| Mouse Kidney (JMS) | 16,473 | 0.836 ± 0.069 | 0.771 ± 0.065 | 0.547 ± 0.071 |
| Mouse Kidney (H&E) | 17,821 | 0.926 ± 0.025 | 0.856 ± 0.042 | 0.656 ± 0.081 |
| Mouse Kidney (PAS) | 21,594 | 0.914 ± 0.039 | 0.814 ± 0.046 | 0.667 ± 0.070 |
| Mouse Kidney (Masson's) | 18,867 | 0.896 ± 0.061 | 0.875 ± 0.067 | 0.743 ± 0.097 |
| Mouse GI (H&E) | 7,768 | 0.887 ± 0.039 | 0.871 ± 0.055 | 0.732 ± 0.098 |
| Mouse GI (PAS) | 28,345 | 0.864 ± 0.075 | 0.806 ± 0.090 | 0.476 ± 0.163 |
| Mouse GI (Masson's) | 12,996 | 0.893 ± 0.067 | 0.837 ± 0.085 | 0.641 ± 0.168 |
| Mouse Brain (H&E) | 8,452 | 0.929 ± 0.058 | 0.848 ± 0.089 | 0.592 ± 0.100 |

Note: GI, gastrointestinal. Metric are computed on 256×256px patches.